\documentclass[aps,prd,twocolumn,secnumarabic,notitlepage,preprintnumbers,superscriptaddress,nofootinbib,tightenlines,floatfix]{revtex4-1}

\usepackage[english]{babel}

\usepackage{graphicx}
\usepackage{amsmath,amssymb}
\usepackage{float}
\usepackage{xcolor}
\usepackage{xfrac}
\usepackage{bm}
\usepackage{ulem}
\usepackage{lineno}
\usepackage[letterpaper,top=2cm,bottom=2cm,left=3cm,right=3cm,marginparwidth=1.75cm]{geometry}

\usepackage{amsmath}
\usepackage{graphicx}
\usepackage[colorlinks=true, allcolors=blue]{hyperref}

\definecolor{hu-berlin-blue}{RGB}{0,65,137} % HEX 004189
\definecolor{hu-berlin-grey}{RGB}{169,169,169}
\definecolor{hu-berlin-red}{RGB}{180,0,0}

\begin{document}

\title{Un-screened forces in Quark-Gluon Plasma ?}

\author{Alexei Bazavov}
\affiliation{Department of Computational Mathematics, Science and Engineering and Department of Physics and Astronomy, Michigan State University, East Lansing, MI 48824, USA}

\author{Daniel Hoying}
\affiliation{Albert Einstein Center, Institute for Theoretical
Physics, University of Bern, CH-3012 Bern, Switzerland}

\author{Rasmus N. Larsen}
\affiliation{Department of Mathematics and Physics, University of Stavanger,
4021 Stavanger, Norway}
\author{Swagato Mukherjee}
\affiliation{Physics Department, Brookhaven National Laboratory, Upton, New York 11973, USA}
\author{Peter Petreczky}
\affiliation{Physics Department, Brookhaven National Laboratory, Upton, New York 11973, USA}
%
%\author{Alexander Karl Rothkopf, Rasmus Normann Larsen, Peter Petreczky, Johannes H. Weber, Alexei Bazavov, Swagato Mukherjee, Olaf Kaczmarek, Daniel Hoying}
\author{Alexander Rothkopf}
\affiliation{Department of Mathematics and Physics, University of Stavanger,
4021 Stavanger, Norway}

\author{Johannes Heinrich Weber}
\affiliation{
Institut f\"ur Physik \& IRIS Adlershof, Humboldt-Universit\"at zu Berlin, D-12489 Berlin, Germany}

\collaboration{HotQCD Collaboration}

\begin{abstract}
We study the correlator of temporal Wilson lines at non-zero temperature in 
2+1 flavor lattice QCD with the aim to define the heavy quark-antiquark potential at 
non-zero temperature. For temperatures $153~{\rm MeV} \leq T \leq 352~{\rm MeV}$ 
the spectral representation of this correlator is consistent with a broadened 
peak in the spectral function, position or width of which then defines the real 
or imaginary parts of the heavy quark-antiquark potential at non-zero temperature, 
respectively. We find that the potential's real part 
%\ELIM{of the potential} 
is not screened contrary to the widely-held expectations. We comment on how this fact may modify 
the picture of quarkonium melting in the quark-gluon plasma.
\end{abstract}

\date{\today}
\preprint{HU-EP-23/48-RTG}

\maketitle

%\section{Introduction}

\section{Introduction}
At very high temperatures the strongly interacting matter undergoes a transition 
to a new state called quark-gluon plasma (QGP). Creating and studying the 
properties of QGP is the goal of large experimental programs in heavy-ion 
collisions at RHIC and LHC \cite{Busza:2018rrf}.

The question of in-medium modifications of the forces between heavy quark $Q$ and 
antiquark $\bar{Q}$ generated a lot of interest since the seminal paper by Matsui and 
Satz \cite{Matsui:1986dk}. They conjectured that color screening in QGP will make 
the $Q\bar Q$ interaction short ranged, and therefore quarkonium states 
cannot be formed in QGP. Thus, QGP formation  in heavy-ion collision will 
lead to quarkonium suppression. The study of quarkonium production
in heavy-ion collisions is a large part of the experimental heavy-ion 
program, see e.g. Ref. \cite{Zhao:2020jqu} for a recent review. 

The idea of having a screened potential between heavy quarks in QGP is closely
related to the exponential screening of the free energy of infinitely heavy 
quarks in QGP, which is well established by lattice QCD calculations, see e.g. 
Ref.~\cite{Bazavov:2020teh} for a review. However, the free energy of heavy 
quarks describes the in-medium interaction of heavy quarks at macroscopic time 
scales much larger than the inverse temperature. 
For understanding the quarkonium properties in QGP one needs to know if and how the 
heavy $Q\bar Q$ potential is modified at scales comparable to the internal 
time scale of quarkonium.
The effective field theory (EFT) approach provides a natural 
framework to address this problem at high temperatures when the weak-coupling 
approach is applicable \cite{Laine:2006ns,Brambilla:2008cx}. Depending on the 
separation of the bound-state scales and the thermal scales the 
heavy $Q\bar Q$ potential can be modified by QGP and also acquire
an imaginary part. In general, however, the real part of this potential does 
not have a screened form in this approach \cite{Brambilla:2008cx}. 
How to study the modification of heavy $Q\bar Q$ interactions in QGP beyond 
weak coupling remains an unsolved problem. However, we could define the
heavy $Q\bar Q$ potential at non-zero temperature ($T>0$) in analogy 
with the zero temperature ($T=0$) case in terms of the Wilson loops of size 
$\tau \times r$ \cite{Rothkopf:2011db}.
We can write the following spectral representation of the 
Wilson loops in terms of the $r$-dependent spectral function
\begin{equation}
    W(\tau,r,T) = \int_{-\infty}^{+\infty} d \omega e^{-\omega \tau} \rho_r(\omega,T).
    \label{eq:spectral}
\end{equation}
The distance $r$ between the 
% \ELIM{infinitely} %
heavy $Q$ and $\bar{Q}$
acts as the label of the spectral function. At $T=0$, the spectral 
function's lowest delta function peak corresponds
to the ground state potential. We expect that there is 
a dominant, broadened peak in the spectral function 
for not too high temperatures; its position and width determine the real and imaginary parts of the potential, respectively~\cite{Rothkopf:2011db}.
For very high temperatures the spectral function may 
lack a well-defined peak such that a potential cannot be defined. 
While the relation between the above defined complex potential and the EFT concept 
of the complex potential is an unsolved problem, too, the existence of a well-defined 
peak in $\rho_r(\omega,T)$ is necessary, yet not a sufficient condition for a 
potential picture of heavy quarkonium at $T>0$.

In this paper we present calculations of the real part of the potential
at $T>0$ in 2+1 flavor QCD using the lattice QCD approach and estimate the imaginary part.
There have been several attempts to
calculate the complex potential at $T>0$ both in 
quenched QCD \cite{Rothkopf:2011db,Bala:2019cqu} as well as in 2+1 flavor QCD \cite{Burnier:2015tda,Bala:2021fkm}. The state of the art calculation of the complex potential in 2+1 flavor QCD has been performed using lattices
with temporal extent $N_{\tau}=12$, and thus at a single lattice
spacing per temperature. The new results are based on 
several lattice spacings and several values of $N_{\tau}$ in the range $N_{\tau}=16-36$. 
The rest of the paper is orgnized as follows. In section \ref{sec:lat} we give some details of
the lattice QCD calculations. Section \ref{sec:anal} presents the analysis
of the lattice results and the main results of the study, while section \ref{sec:concl} contains our conclusions. Many technical details of the calculations
are discussed in the Appendix.
\begin{figure}
\includegraphics[width=0.45\textwidth]{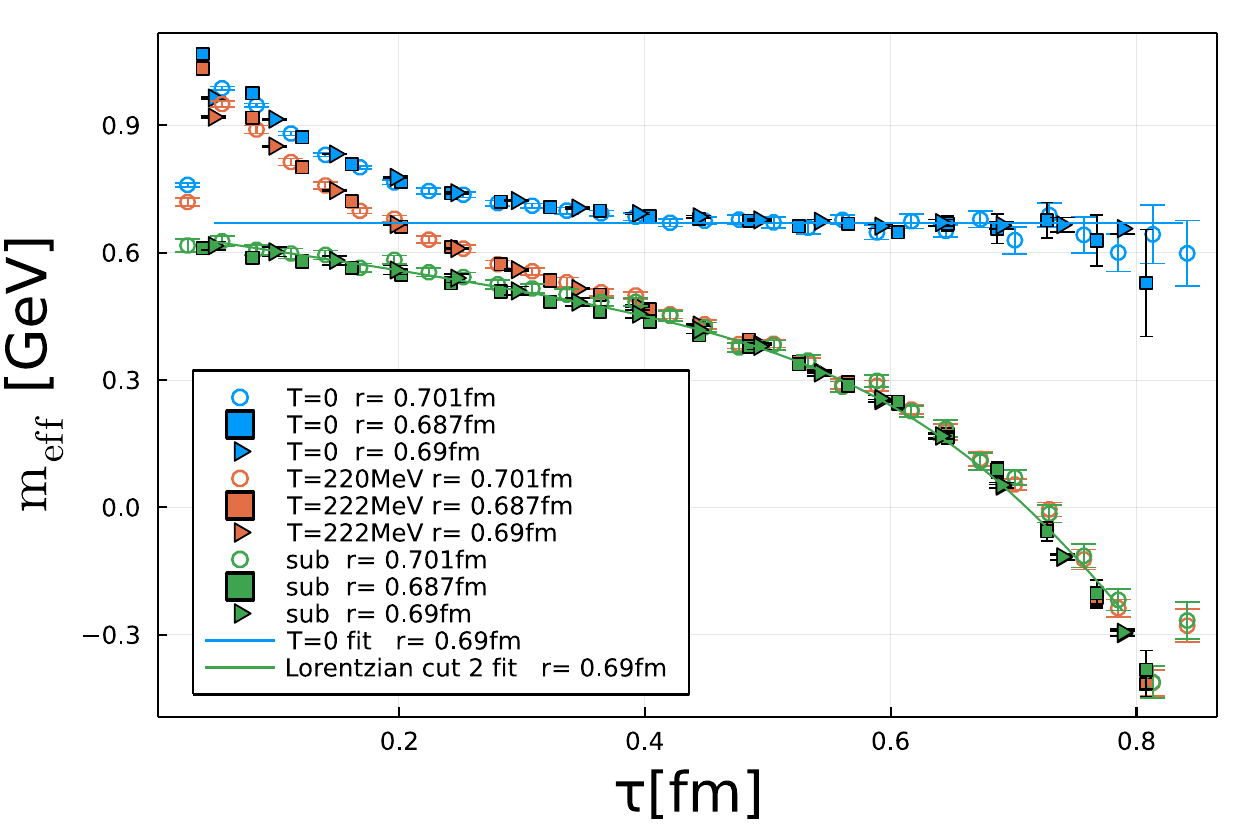}
%\vspace*{-1.4cm}
\caption{
The effective masses at $T=0$ and at $T \simeq 220$ MeV for $r \simeq 0.7$ fm and  
$a=0.0280$ fm (circles), $a=0.0404$ fm (squares) 
or 
$a=0.0493$ fm (triangles). 
The green 
symbols correspond to 
subtracted data. 
The lines show the fits discussed in the text. 
Filled (open) symbols represent $m_s/m_l=$20(5).}
\label{fig:meff_sub}
%\vspace*{-0.3cm}
\end{figure}

\section{Details of the lattice QCD calculations}
\label{sec:lat}
In lattice QCD 
one often considers correlators of Wilson lines in 
Coulomb gauge instead of Wilson loops since these contain
the same physical information and are less noisy, 
see the discussions in Ref. \cite{Bala:2021fkm} and Appendix \ref{app:lat}. 
We calculated the 
Wilson line correlators $W(\tau,r,T)$ in 2+1 flavor QCD 
using highly improved 
staggered quark (HISQ) action \cite{Follana:2006rc} and 
L\"uscher-Weisz
action \cite{Luscher:1984xn,Luscher:1985zq} on $N_s^3 \times N_{\tau}$ lattices for physical strange quark mass, $m_s$ and 
two sets of light ($u$ and $d$) quark mass, $m_l=m_s/5$ and $m_l=m_s/20$. The latter 
corresponds to almost physical pion mass, $m_{\pi}=161$ MeV in the continuum limit. 
Furthermore, the calculations have been performed for three different lattice spacings
corresponding to the following values of bare lattice gauge coupling $\beta=10/g_0^2=7.596,~7.825$
and $8.249$. 
The lattice spacing and thus the temperature scale $T=\sfrac{1}{(aN_\tau)}$ has been fixed using 
the $r_1$-scale determined in Ref.~\cite{Bazavov:2017dsy} with the value $r_1=0.3106\,\mathrm{fm}$ 
obtained in Ref.~\cite{MILC:2010hzw}. The value of the strange quark mass was obtained from the 
parametrization of the line of constant physics from Ref.~\cite{HotQCD:2014kol}. 
We this scale setting for the lattice spacing we obtain: $a(\beta=8.249)=0.0280$ fm, 
$a(\beta=7.285)=0.0404$ fm and $a(\beta=7.596)=0.0493$ fm.
For the finest lattices the spatial size of the lattice is $N_s=96$, while for the
two coarser lattices we use $N_s=64$. The temporal size of the lattice is varied 
in the range $N_{\tau}=16-36$, which corresponds to the temperature 
range $153~{\rm MeV}\le T \le 352$ MeV. Further details about the parameters of
the lattice calculations are given in Appendix \ref{app:lat}.

For the smallest lattice spacing, we only consider $T\ge 195$ MeV, i.e. temperatures
well above the chiral crossover temperatures. For these temperatures we do not
expect significant quark mass dependence of the Wilson line
correlators. Therefore, the calculations for the smallest lattice
spacing have been performed only with $m_l=m_s/5$, while for 
the coarser lattices we use $m_l=m_s/20$.
As discussed later,
we do not see any $m_l$ dependence of the Wilson line correlator for $T\ge 195$ MeV, as expected.

For lattices with large temporal extents, employed in this study, noise reduction methods have to be used. We
use gradient flow \cite{Luscher:2010iy} for noise reduction. 
To reduce noise even further, we require that $W(\tau,r,T)$ at 
large $r/a$ is a smooth function of $r/a$, and replace it for each value 
$\tau$ by a corresponding local $r/a$ interpolation.
We verified that this procedure does not introduce bias in
our analysis by varying the interpolation intervals and
comparing to the results that do not use interpolation.
Further details on the noise reduction techniques are presented Appendix \ref{app:lat}.

To aid the reconstruction of 
the spectral function we also performed calculations on $N_{\tau}=64$ and 
$N_{\tau}=56$ lattices, which we refer to as $T=0$ lattices. 
% \begin{figure}[H]
% % \includegraphics[width=0.5\textwidth]{b8249_sub_meff_r10_Nt20.pdf}\\
% %\includegraphics[width=0.5\textwidth]{b8249_sub_meff_r14_Nt28.pdf}\\
% %\includegraphics[width=0.5\textwidth]{b7825_sub_meff_r16_Nt26.pdf}
% \includegraphics[width=0.5\textwidth]{meff_lattice_comp_T220_r25_updated.pdf}
% \caption{
% %The subtracted effective masses as a function of $\tau$ at $T=352$ MeV, $r=0.28$ fm (top) and 
% The effective masses at $T=0$ and at $T \simeq 220$ MeV for $r \simeq 0.7$ fm and 
% % three different lattice spacings, 
% $a=0.0280$ fm (circles), $a=0.0404$ fm (squares) 
% % and
% or 
% $a=0.0493$ fm (triangles). 
% The green 
% symbols correspond to 
% % the subtracted effective masses. 
% subtracted data. 
% The lines show the fits discussed in the text. 
% \RESUB{Filled (open) symbols represent $m_s/m_l=$20(5).}
% }
% \label{fig:meff_sub}
% \end{figure}

\section{Analysis and Results} 
\label{sec:anal}
To analyze the 
% \ELIM{lattice results on the} 
Wilson line correlator $W(\tau,r,T)$ 
in Eq. (\ref{eq:spectral}) it is useful to consider the effective mass defined as 
\begin{align}
    m_{\text{eff}}(\tau,r,T)  
    &= -\partial _\tau \ln W(\tau,r,T) 
    \nonumber \\
    &
    = -\frac{1}{a}\ln\Bigg[\frac{W(\tau+a,r,T)}{W(\tau,r,T)}\Bigg]~,
    \label{eq:meff}
\end{align}
where the last equation applies to the case of non-zero lattice spacing. 
At $T=0$, the effective mass decreases
with increasing $\tau$, and reaches a plateau
for sufficiently large $\tau$%
, since the 
spectral function is positive definite and has the lowest 
ground state delta function peak followed by many excited states for $\omega$ above 
the ground state.
We show the results for the effective masses  in Fig. \ref{fig:meff_sub}. 
We see that at $T=0$ the effective mass decreases with increasing $\tau$, with 
the exception of the data at smallest $\tau$, and approaches a
plateau for $\tau$ around $0.5$ fm. 
The non-monotonic behavior 
is due to the smearing artifacts coming from the gradient flow,  as
discussed in Appendix \ref{app:anal0}.
Except for very small $\tau$,
$m_{\text{eff}}$ decreases at $T>0$ with increasing $\tau$ for all 
$\tau$ values and does not reach a plateau. 
This means that there is no stable ground state at non-zero temperature. 
We see from Fig. \ref{fig:meff_sub} that the effective masses show neither 
lattice spacing nor sea quark mass dependence for $T>200$ MeV. This implies
that for these temperatures using $m_l=m_s/5$ is equivalent to using the physical
light quark mass and that our results are essentially in the continuum limit.
We also compared the effective masses corresponding to different lattice
spacings at lower temperatures and found no dependence on the lattice spacing.

At small $\tau$ the difference between the $T=0$ or $T>0$
effective masses is the smallest, and their $\tau$-dependence
is rather similar,
see Fig. \ref{fig:meff_sub}. Thus, we aim
to constrain the corresponding contributions at $T>0$ by using the $T=0$ results.

Our objective is to extract information on a dominant peak in
the spectral function corresponding to 
$W(\tau,r,T)$ at $T>0$. 
We choose an Ansatz \cite{Bala:2021fkm} for the spectral function as
\begin{equation}
    \rho_r(\omega,T)=
    \rho_r^{\text{low}}(\omega,T)+\rho_r^{\text{peak}}(\omega,T)+\rho_r^{\text{high}}(\omega),
    \label{eq:ansatz}
\end{equation}
where $\rho_r^{\text{high}}(\omega)$ is assumed to be a temperature-independent part 
dominating at large $\omega$. 
$\rho_r^{\text{peak}}(\omega,T)$ 
describes a dominant peak encoding the complex potential at $T>0$, 
while $\rho_r^{\text{low}}(\omega,T)$ is a small, medium-dependent contribution 
below the dominant peak. 

Fixing $\rho_r^{\text{high}}(\omega)$ to its $T=0$ value 
effectively means subtracting it from the $T>0$
result. We define the subtracted correlator as follows,
\begin{equation}
    W^{\text{sub}}(\tau,r,T)=
    W(\tau,r,T)-W^{\text{high}}(\tau,r),
\end{equation}
which is solely determined by the medium-dependent 
contributions to the spectral function, 
\begin{equation}
\rho_r^{\text{low}}(\omega,T)+\rho_r^{\text{peak}}(\omega,T)
=\rho_r(\omega,T)-\rho_r^{\rm high}(\omega).
\end{equation}
Since the $T=0$ spectral function has the form
\begin{equation}
    \rho_r(\omega,T=0)=A_r\delta(\omega-V(r,T=0))+\rho_r^{\text{high}}(\omega),
\end{equation}
we define 
\begin{equation}
W^{\text{high}}(\tau,r) \equiv \int_{-\infty}^{\infty} d \omega \rho^{\text{high}}_r(\omega) e^{-\omega \tau}
\end{equation}
via 
\begin{equation}
    W^{\text{high}}(\tau,r)=W(\tau,r,T=0)-A_r e^{-V(r,T=0) \tau}.
    \label{eq:def Whigh}
\end{equation}
Thus, it is straightforward to estimate $ W^{\text{high}}(\tau,r)$ 
using single-exponential fits for $A_r$ and $V(r,T=0)$.
The task of constraining $\rho_r^{\text{peak}}(\omega,T)$
and $\rho_r^{\text{low}}(\omega,T)$ is now reduced to the analysis of the $\tau$-dependence of $W^{\text{sub}}(\tau,r,T)$.

The effective masses from $W^{\text{sub}}(\tau,r,T)$ at \mbox{$T>0$}
are also shown in Fig. \ref{fig:meff_sub}. The uncertainties
in the effective masses due to the errors in the determination
of the ground state contribution at $T=0$ have been taken into
account by combining these uncertainties with the statistical
errors of the $T>0$ calculations.
We note that the non-monotonic behavior at small
$\tau$ due to smearing artifacts is absent in these subtracted effective 
masses $m_{\text{eff}}^{\text{sub}}(\tau,r,T)$,
and therefore, these artifacts do not 
affect $\rho^{\text{peak}}_r(\omega,T)$, see Appendix \ref{app:analT}.
As discussed in Appendix \ref{app:analT}, $m_{\text{eff}}^{\text{sub}}(\tau,r,T)$ would 
decrease linearly in $\tau$ if the ground state peak had a Gaussian shape \cite{Bala:2021fkm}.
$m_{\text{eff}}^{\text{sub}}(\tau,r,T)$ 
shows linear behavior in $\tau$ 
at small $\tau$, indicating that the dominant 
ground state peak has broadened. 
Here we note, that the behavior of the subtracted effective masses
obtained from the Wilson line correlators and from the Wilson loops is the same \cite{Bala:2021fkm}.

As discussed in Ref. \cite{Bala:2021fkm} 
$\rho_r^{\text{low}}(\omega,T)$ is the contribution to the spectral
function at $T>0$ which has support for energies well below the dominant peak and
representing a heavy $Q\bar Q$ state propagating forward in Euclidean
time interacting with a backward propagating light state from the medium. 
This contribution is much smaller than $\rho_r^{\text{peak}}(\omega,T)$ but 
dominates the correlator at $\tau$ around $1/T$. This part of the spectral function 
explains the rapid drop of $m_{\text{eff}}(\tau,r,T)$ at large $\tau$ \cite{Bala:2021fkm} that 
can be seen in Fig. \ref{fig:meff_sub}.

A physically appealing parametrization of $\rho^{\text{peak}}(\omega,T)$ is a Lorentzian form. 
However, a Lorentzian form is only valid in the vicinity of the peak. In general, we can assume that the correlator has a pole at some complex $\omega$, so 
\begin{align}
\rho_r^{\text{peak}}(\omega,T) 
&= \frac{1}{\pi}{\rm Im} \frac{A_r(T)}{\omega-{\rm Re} V(r,T)-i \Gamma(\omega,r,T)}.
\end{align}
For $\omega \simeq {\rm Re} V(r,T)$ we can approximate $\Gamma(\omega,r,T)$ by a constant: 
$\Gamma(\omega,r,T) \simeq \Gamma_L(r,T)$.
However, for $\omega$ values far away from the peak $\Gamma(\omega,r,T)$ must
quickly go to zero. The self-consistent $T$-matrix calculation of heavy $Q\bar Q$ 
propagators indeed shows an exponential decrease of $\Gamma(\omega,r,T)$ 
away from the peak \cite{Liu:2017qah}. To incorporate this feature of the 
spectral function in our analysis we assume that $\rho_r^{\text{peak}}(\omega,T)$ is given by 
$\Gamma_L(r,T)/([\omega-{\rm Re}V(r,T)]^2+\Gamma_L^2(r,T))$ for 
\mbox{$|\omega-{\rm Re} V(r,T)| \lesssim \Gamma_L(r,T)$} and is zero otherwise. 
Such a cut Lorentzian form 
gives rise to an almost linear behavior of $m_{\text{eff}}^{\text{sub}}(\tau,r,T)$ at small $\tau$, too, 
as required by the lattice data.
\begin{figure}[H]
\centering
\includegraphics[width=0.45\textwidth]{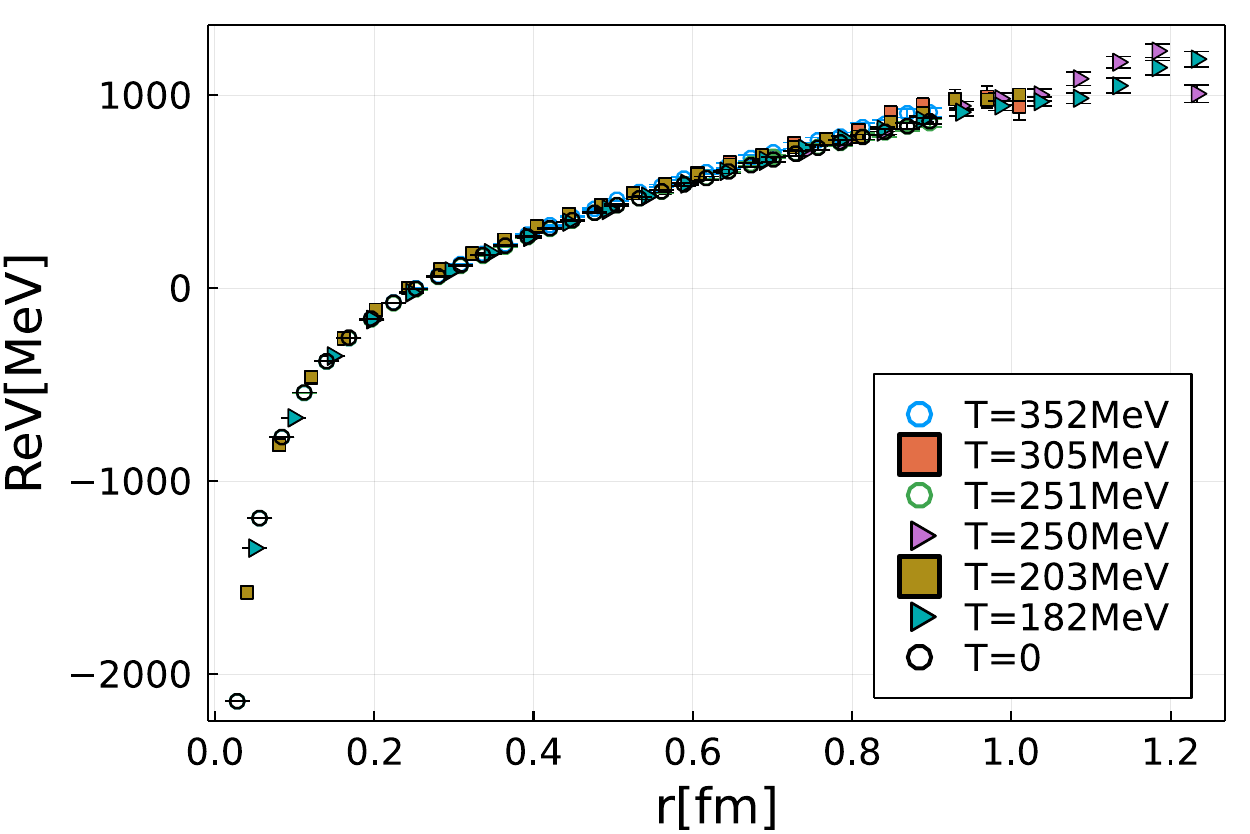}
\caption{The real part of the potential as a function of $r$ at different temperatures. We show results for 
$a=0.0280$ fm (circles), $a=0.0404$ fm (squares) 
or 
$a=0.0493$ fm (triangles). Open symbols for $m_s/m_l=5$ and filled symbols for $m_s/m_l=20$.}
\label{fig:ReV}
\end{figure}
\begin{figure*}
\centering
\includegraphics[width=0.32\textwidth]{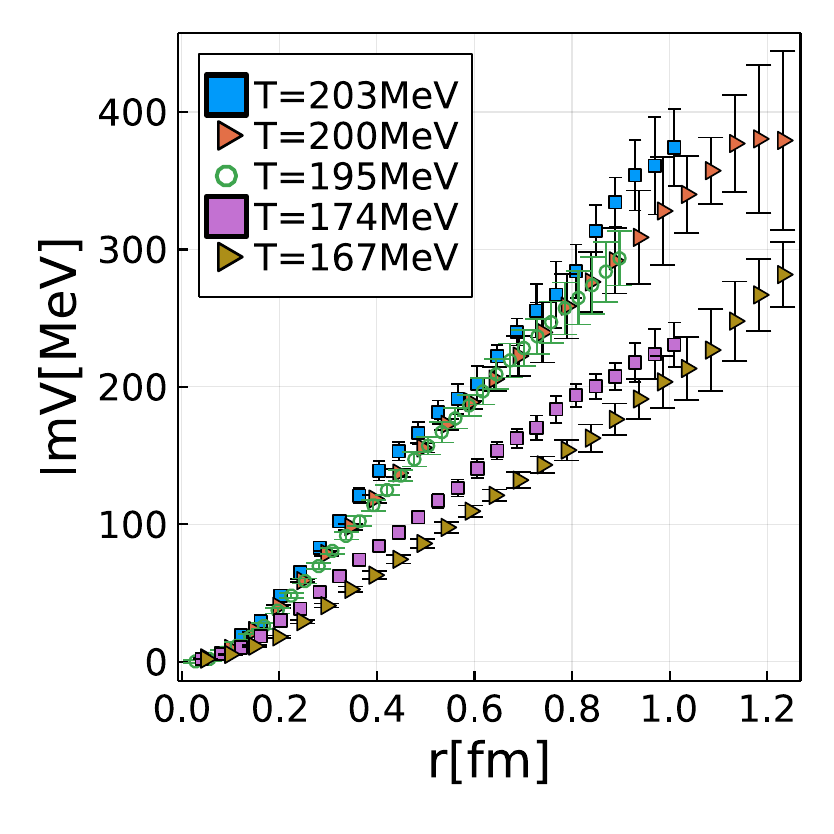}
\includegraphics[width=0.32\textwidth]{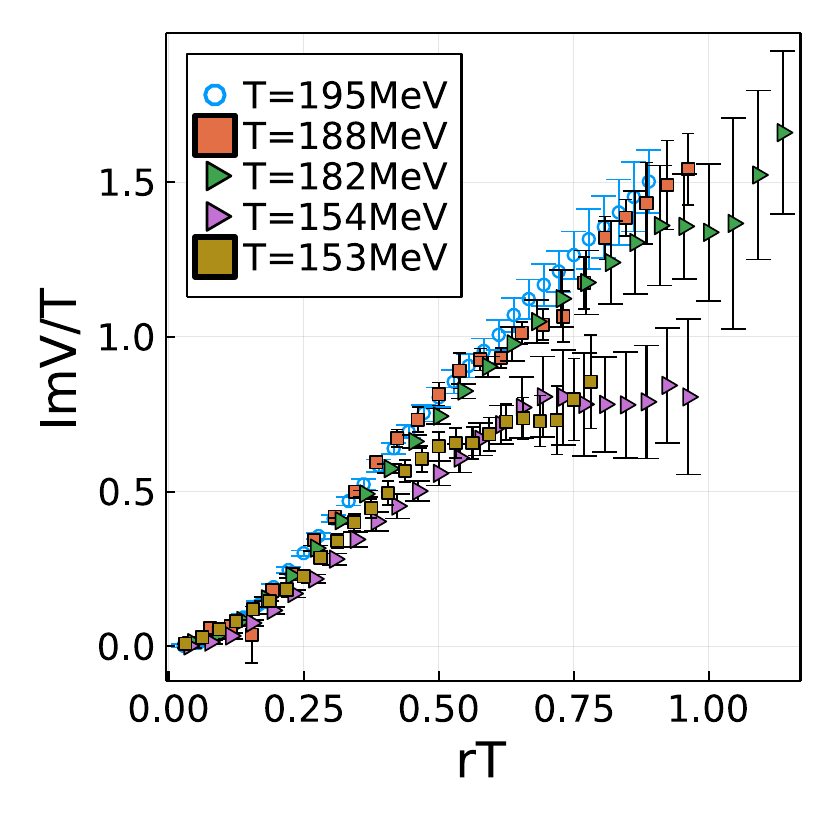}
\includegraphics[width=0.32\textwidth]{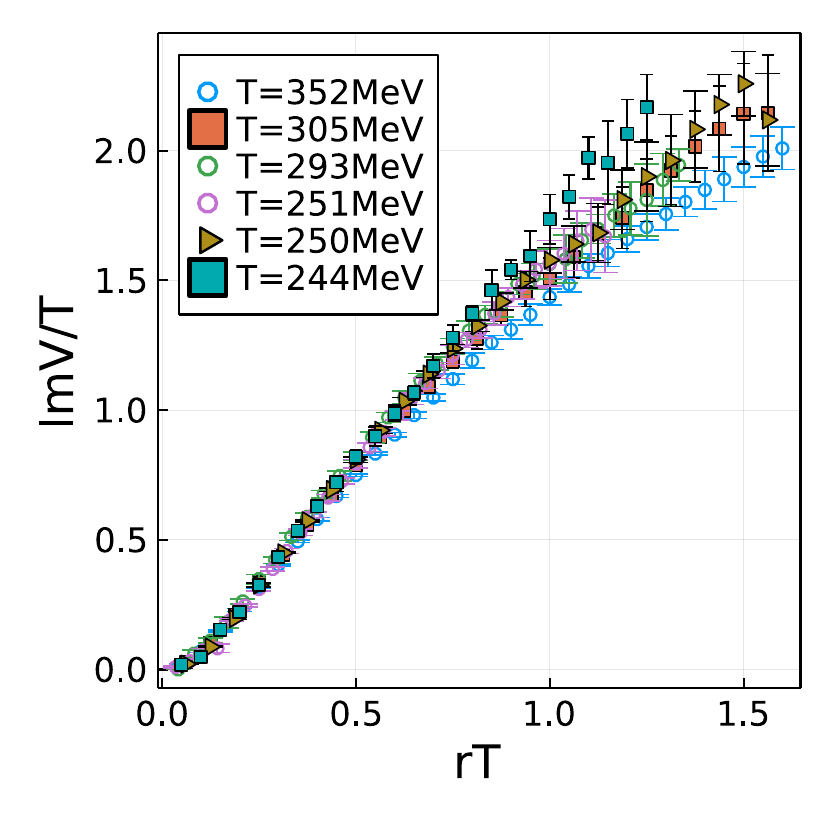}
\caption{
%\vspace*{0.3cm}
The estimate of the imaginary part of the potential from the fit using 
cut Lorentzian form as a function of $r$ or $rT$ for different temperatures. 
The three panels focus on different temperature ranges. 
The circles correspond to 
$a=0.0280$ fm, the squares to 
$a=0.0404$ fm, and the triangles 
correspond to 
$a=0.0493$ fm. Open symbols for $m_s/m_l=5$ and filled symbols for $m_s/m_l=20$.
Error bars include a systematic contribution discussed Appendix \ref{app:cum}.}
\label{fig:ImV}
\end{figure*}
\vspace*{-0.5cm}
The most general parametrization of $\rho_r^{\text{low}}(\omega,T)$ would be a sum of delta 
functions at $\omega$ well below the dominant peak position. However, to describe our
effective mass data even 
a single delta function at sufficiently small $\omega$, 
$\rho_r^{\text{low}}(\omega,T)=c_r^{\text{low}}(T) \delta(\omega-\omega_r^{\text{low}}(T))$%
~turns out as sufficient%
.

With these forms of $\rho_r^{\text{peak}}(\omega,T)$ and $\rho_r^{\text{low}}(\omega,T)$ we 
fitted the lattice data on $m_{\text{eff}}^{\text{sub}}(\tau,r,T)$
and determined the fit parameters ${\rm Re}V(r,T)$, 
$\Gamma_L(r,T)$, $c_r^{\text{low}}(T)/A_r(T)$ and $\omega_r^{\text{low}}(T)$. 
A sample fit is shown in Fig. \ref{fig:meff_sub} and
details of the fits are discussed in Appendix \ref{app:analT}.
We typically find that $c_r^{\text{low}}(T)/A_r(T)<5 \cdot 10^{-4}$ and decreases
with decreasing $r$, while $\omega_r^{\text{low}}(T)$ is between (1.8-3.8) GeV
below the peak position $\omega={\rm Re} V(r,T)$.

The results for ${\rm Re}V(r,T)$ are shown in Fig. \ref{fig:ReV}
indicating a temperature-independent real part in good agreement with the $T=0$ potential. 
This is not completely unexpected, as 
$m_{\text{eff}}(\tau,r,T)$
at small $\tau$ is close to the vacuum result, c.f. Fig. \ref{fig:meff_sub}. 
The peak position is insensitive to the detailed shape of $\rho_r^{\text{peak}}(\omega,T)$,
i.e. for 
a Gaussian form we find the same peak position within errors. 
Thus our lattice QCD results show that the potential's real part 
is unscreened. 
This observation supersedes conclusions drawn earlier by applying the Bayesian 
Reconstruction (BR) method~\cite{Burnier:2013nla} to older lattice data~\cite{Burnier:2014ssa} 
with much larger statistical errors and larger discretization artifacts. 
There are distortions in ${\rm Re} V(r,T)$ at the two shortest
distances in lattice units ($r=a, 2a$), but these distortions are
the same both at $T=0$ or $T>0$, see the discussion in Appendix \ref{app:anal0} and Appendix \ref{app:analT},
and do no affect
our conclusion about the absence of screening.

As discussed above the imaginary part of the potential is defined as the width of 
the ground state peak at $T>0$.
If we knew 
the spectral function exactly we could
fit it 
in the peak's vicinity 
with a Lorentzian form, whose width parameter
would give the potential's imaginary part. 
This has been 
explicitly checked for the spectral function of an infinitely heavy $Q\bar Q$ pair
calculated in hard thermal loop perturbation theory \cite{Burnier:2013fca}.
Yet the correlator is sensitive to all parts of the spectral 
function, in particular to $\rho_r^{\text{low}}(\omega,T)$ and to the tails of $\rho_r^{\text{peak}}(\omega,T)$.
For this reason, the parameter $\Gamma_L$ cannot be considered as ${\rm Im} V(r,T)$. 
A better way to characterize ${\rm Im} V(r,T)$ is to consider the cumulants of 
$\rho_r^{\text{peak}}(\omega,T)$. 
The first two cumulants are defined as $c_1=\langle \omega \rangle$
and $c_2=\langle \omega ^2\rangle-\langle \omega \rangle ^2$, where 
$\langle \dots \rangle = \int d \omega \dots$. In the case of the Gaussian, the 
second cumulant of the spectral function is the square of the width parameter. In the case of 
the cut Lorentzian, it is proportional to the square of the parameter $\Gamma_L$. 
Furthermore, if $c_r^{\text{low}}/A_r$ is very small, $\rho_r^{\text{peak}}(\omega,T)$ determines
the behavior of the Wilson line correlator around $\tau=0$. 
Therefore, the second cumulant of $\rho_r^{\text{peak}}(\omega,T)$ 
determines the slope of $m_{\text{eff}}^{\text{sub}}(\tau,r,T)$
at small $\tau$, which is well defined from the lattice data, see the Appendix \ref{app:cum}.
Thus the square root of the second cumulant of $\rho_r^{\text{peak}}(\omega,T)$ is a good proxy for the $r$ and 
temperature dependence of ${\rm Im} V(r,T)$. 
In Fig. \ref{fig:ImV} we show this proxy for ${\rm Im} V(r,T)$ as a function of distance $r$ 
for different temperatures.
We scaled the $x$- and $y$-axes by the temperature in the two middle and right panels of Fig. \ref{fig:ImV}.
We see that for $180~{\rm MeV}< T \le 352$ MeV the numerical results for ${\rm Im} V(r,T)$ scale with 
the temperature, i.e. the imaginary part of the potential depends only on $rT$ and is proportional to
the temperature. This is in qualitative agreement with the weak-coupling results. 
Since for $rT \simeq 1$ the imaginary
part of the potential is larger than the temperature, the forces between heavy quarks 
are damped very quickly, i.e. on the time scale comparable to or shorter than the 
thermal scale. During that short time scale, the chromo-electric field between the 
heavy $Q$ and $\bar Q$ cannot adjust itself to the medium. 
The chromo-electric force between the heavy quarks is simply damped 
away, and the heavy $Q$ and $\bar Q$ will not interact. This picture of quarkonium
melting is very different 
from the one proposed by Matsui and Satz.
While ${\rm Im} V$ is quite large we still think the $Q\bar Q$ energy
is well defined in the considered temperature
interval. For if there would be no well defined dominant peak
in the spectral functions, different static $Q\bar Q$ correlators
would have quite different $\tau$ dependence. However, as shown
in our previous study \cite{Bala:2021fkm} this is not the case.

\section{Conclusion} 
\label{sec:concl}

We studied the complex heavy quark-antiquark potential at 
non-zero temperature in 2+1 flavor QCD using lattice calculations with a large 
temporal extent. We have found that contrary to some common expectations the real 
part of the potential is not screened for temperatures $153~{\rm MeV} \le T \le 352$ MeV. 
We also found that the dissipative effects on the chromo-electric forces between 
the heavy quarks, encoded in the imaginary part of the potential are very large
and likely will lead to quarkonium dissolution.

As already mentioned in the introduction, the lack of screening in the real part of the potential is expected
in weak-coupling limit for $r T<1$ \cite{Brambilla:2008cx}. Our study
shows that this also holds non-perturbatively. Furthermore, a numerical 
evaluation of the weak-coupling result of the thermal correction to the real part of the potential shows that this correction is quite small. For $T>500$
MeV and $rT<0.4$, where the weak coupling result of Ref. \cite{Brambilla:2008cx}
may be applicable, the thermal correction to the real part of the potential is smaller 
than $0.1 T$.

\section*{Acknowledgements}
The computations in this work were performed using \texttt{SIMULATeQCD}~\cite{Mazur:2023lvn} and the MILC code\footnote{For details see \href{https://github.com/milc-qcd/}{MILC's GitHub page}}.

R.~L. and A.~R. acknowledge support by the Research Council of Norway under the FRIPRO Young Research Talent grant 286883.

This material is based upon work supported by The U.S. Department of Energy, Office of Science, Office of Nuclear Physics through Contract No.~DE-SC0012704, and within the frameworks of Scientific Discovery through Advanced Computing (SciDAC) award Fundamental Nuclear Physics at the Exascale and Beyond and the Topical Collaboration in Nuclear Theory Heavy-Flavor Theory (HEFTY) for QCD Matter. 
A.B.'s research is supported by the U.S. National Science
Foundation under award PHY-1812332.
J.H.W.’s research is funded by the Deutsche Forschungsgemeinschaft (DFG, German Research
Foundation)---Projektnummer 417533893/GRK2575 ``Rethinking Quantum Field Theory''.

This research used awards of computer time provided by the National Energy Research Scientific Computing Center (NERSC), a U.S. Department of Energy Office of Science User Facility located at Lawrence Berkeley National Laboratory, operated under Contract No. DE-AC02- 05CH11231, and the PRACE awards on JUWELS at GCS@FZJ, Germany and Marconi100 at CINECA, Italy. Computations for this work were carried out in part on facilities of the USQCD Collaboration, which are funded by the Office of Science of the U.S. Department of Energy.

\bibliography{ref.bib}

\appendix

\section{Lattice QCD setup}
\label{app:lat}

In this appendix we discuss further details of our lattice QCD calculations.
The parameters 
of the lattice calculations including the lattice volume and the quark masses are given in 
Tables \ref{table:data_set_8.249}, \ref{table:data_set_7.825}, and \ref{table:data_set_7.596}.
The gauge configurations used in this study have been generated using a rational hybrid Monte-Carlo 
algorithm \cite{Clark:2003na} with grants from PRACE on Juwels Booster and Marconi 100 and NERSC on 
Perlmutter using the SIMULATeQCD code \cite{Mazur:2023lvn}.
We also used the MILC code 
on Cori at NERSC to generate the gauge configurations. Some of the gauge configurations have 
been generated on the USQCD cluster in JLab. After removing the initial trajectories for thermalization 
we arrived at the data set in Tables \ref{table:data_set_8.249}, \ref{table:data_set_7.825}, and 
\ref{table:data_set_7.596}. Every 5th trajectory has been used for $N_\sigma=96$ and every 10th trajectory 
for $N_\sigma=64$.

On the generated gauge configurations we calculated Wilson line correlators in Coulomb gauge 
with the aim of determining the static quark-antiquark ($Q\bar Q$) potential. We use Wilson line 
correlators instead of Wilson loops because these are much less noisy and provide more 
convenient access to distances at non-integer multiples of the lattice spacing. 
At $T=0$ both Wilson loops and Wilson line correlators in Coulomb gauge have been used for the 
determination of the $Q\bar Q$ potential, see e.g. Refs. \cite{Aubin:2004wf, Cheng:2007jq, 
Bazavov:2011nk, Bazavov:2017dsy, Brambilla:2022het}. In the case of Wilson loops, smearing should 
be applied to the spatial gauge links entering the Wilson loops in order to obtain a reasonable 
signal. In Ref. \cite{Bala:2021fkm} both Wilson lines and Wilson loops with three-dimensional
hyper-cubic (HYP) smearing \cite{Hasenfratz:2001hp} in the spatial gauge links have been studied 
at non-zero temperature. It was found there that the behavior of the Wilson line correlators and 
Wilson loops is fairly similar except for small $\tau$, where sensitivity to 
excited states is different, similar to the $T=0$ case~\cite{Bazavov:2019qoo}. 
At $T>0$ there are also some differences between the behavior of Wilson loops and
Wilson line correlators at $\tau \simeq 1/T$, which are, however, not related to
$Q\bar Q$ potential as discussed below.
Thus both Wilson lines in Coulomb gauge and Wilson loops encode the same temperature modification 
of the $Q \bar Q$ potential. In Ref. \cite{Bala:2021fkm} the calculations of the Wilson lines have 
been performed on $N_{\tau}=12$ lattices. Since we use much larger $N_{\tau}$ in this study, also 
the temporal links have to be smeared. We use gradient flow \cite{Luscher:2010iy} for the smearing 
of the temporal gauge links. More precisely we use Zeuthen flow \cite{Ramos:2015baa}. For flow time 
$\tau_F$ the gauge links are smeared in a radius $\sqrt{8 \tau_F}$. This radius should be much 
smaller than the inverse temperature. We use different flow times corresponding to the flow radius 
in the range $a-2.53a$ and study the sensitivity of our results to the flow time. For the final 
results presented in the paper, we use the smallest flow time that gives an acceptable signal. Since 
the signal deteriorates with increasing $N_{\tau}$ we use larger flow time for large $N_{\tau}$. 
The range of flow times and the specific values of flow times for which we show the final result 
are presented in Tables \ref{table:data_set_8.249}, \ref{table:data_set_7.825}, and 
\ref{table:data_set_7.596} for $\beta=8.249,~7.825$ and $7.596$, respectively. 

After performing the gradient flow we fix the Coulomb gauge. The precision of Coulomb gauge fixing 
was set to $10^{-6}$. We also note that neither is the gradient flow the only option to smear 
the temporal gauge links nor is it a problem to fix the Coulomb gauge before performing the gradient 
flow, when studying Wilson line correlators. Previously we used HYP smearing after gauge fixing for 
the temporal gauge links when calculating the Wilson line correlators at $T>0$ \cite{Hoying:2021mba} 
and found that the temperature and the $\tau$ dependence of the correlators are similar to that 
reported here. Thus even though smearing destroys the
gauge fixing condition to some extent, the qualitative behavior of
the Wilson line correlators is not affected.
%\RNL{Wilson smearing destroys the gauge fixing slightly so the order is important.}
%\JHW{Yes it is a different gauge, but the question what is the exact gauge doesn't matter as long as it 
%answered with always the same. The states are the same and the overlap factors change very little}
This implies that our findings are neither sensitive to the details of gauge link smearing nor 
to details of the Coulomb gauge fixing.
\begin{table}[h!]
\begin{center}
\begin{tabular}{||c c c c c ||} 
 \hline
$N_\tau$ & $\#$ & $m_s/m_l$ & $T[MeV]$ & $\tau_F/a^2$ \\ [0.5ex] 
 \hline\hline
 20 & 3200 & 5 & 352 & 0.125 \\ 
 \hline
 24 & 856 & 5 & 293 & 0.125 \\
 \hline
 28 & 2400 & 5 & 251 & 0.2\\
 \hline
 32 & 1100 & 5 & 220 & 0.4 \\
 \hline
 36 & 2400 & 5 & 195 & 0.6 \\
  \hline
 56 & 1000 & 5 & 126 & 0.125,0.2,0.4,0.6 \\ [1ex] 
 \hline 
\end{tabular}
\caption{Parameters for the $N_\sigma=96$, $\beta = 8.249$, 
$am_s=0.01011$ lattice configurations used. The last column 
shows the flow time used for each $N_{\tau}$.
}
\label{table:data_set_8.249}
\end{center}
\end{table}

\begin{table}[h!]
\begin{center}
\begin{tabular}{||c c c c c ||} 
 \hline
$N_\tau$ & $\#$ & $m_s/m_l$ & $T[MeV]$ &  $\tau_F/a^2$ \\ [0.5ex] 
 \hline\hline
 16 & 5528 & 20 & 305 & 0.0-0.6 [0.125] \\ 
 \hline
 18 & 5230 & 20 & 271 & 0.0-0.6 [0.125] \\
 \hline
 20 & 4726 & 20 & 244 & 0.0-0.6 [0.125]\\
 \hline
 22 & 3515 & 20 & 222 & 0.0-0.6 [0.125]\\
 \hline
 24 & 3345 & 20 & 203 & 0.0-0.6 [0.2]\\
  \hline
 26 & 4147 & 20 & 188 & 0.0-0.6 [0.2]\\ [1ex] 
 \hline
 28 & 3360 & 20 & 174 & 0.0-0.6 [0.4]\\ [1ex] 
 \hline 
 30 & 2679 & 20 & 163 & 0.0-0.6 [0.4]\\ [1ex] 
 \hline 
 32 & 2133 & 20 & 153 & 0.0-0.6 [0.6]\\ [1ex] 
 \hline 
 64 & 1006 & 20 & 76 & 0.0-0.6 [0.125-0.6]  \\ [1ex] 
 \hline 
\end{tabular}
\caption{Parameters for $N_\sigma=64$, $\beta = 7.825$, $am_s=0.0164$ lattice configurations.
The last column shows the range of flow time in lattice units used in the 
calculations. The numbers in the square brackets indicate the flow time for
which the final results in the paper are presented.}
\label{table:data_set_7.825}
\end{center}
\end{table}

\begin{table}[h!]
\begin{center}
\begin{tabular}{||c c c c c ||} 
 \hline
$N_\tau$ & $\#$ & $m_s/m_l$ & $T[MeV]$ & $\tau_F/a^2$ \\ [0.5ex] 
 \hline\hline
 16 & 4697 & 20 & 250 & 0.0-0.8 [0.2]\\ 
 \hline
 18 & 3715 & 20 & 222 & 0.0-0.8 [0.2] \\
 \hline
 20 & 3005 & 20 & 200 & 0.0-0.8 [0.4]\\
 \hline
 22 & 4158 & 20 & 182 & 0.0-0.8 [0.4]\\
 \hline
 24 & 3278 & 20 & 167 & 0.0-0.8 [0.6]\\
  \hline
 26 & 2423 & 20 & 154 & 0.0-0.8 [0.8]\\ [1ex] 
% \hline
% 28 & 2030 & 20 & 143 & 0.0-0.8 & 0.8\\ [1ex] 
% \hline 
% 30 & 1677 & 20 & 133 & 0.0-0.8 & 0.8\\ [1ex] 
 \hline 
 64 & 914 & 20 & 63 & 0.0-0.8 [0.2-0.8]\\ [1ex] 
 \hline 
\end{tabular}
\caption{Parameters  for $N_\sigma=64$, $\beta = 7.596$, $am_s=0.0202$ lattice configurations.
The last column shows the range of flow time in lattice units used in the 
calculations. The numbers in the square brackets indicate the flow time for
which the final results in the paper are presented.}
\label{table:data_set_7.596}
\end{center}
\end{table}

\section{Analysis of the Wilson line correlators at $T=0$}
\label{app:anal0}

In this appendix we discuss the analysis of the Wilson line correlators at zero
temperature.
For the analysis of the Wilson line correlators, it is useful to consider the effective
masses defined in Eq. (\ref{eq:meff}.
The Wilson line correlators require
multiplicative renormalization which corresponds to an additive normalization of the effective
masses that is proportional to $1/a$. This normalization can be fixed by requiring for each 
lattice spacing that the $Q\bar Q$ potential at $T=0$ is equal to a prescribed value for 
one given distance. Here we use the prescription $V(r=r_0)=0.954/r_0$, where $r_0$ is the 
Sommer scale, which for 2+1 flavor QCD is $r_0=0.468(4)$ fm \cite{Bazavov:2011nk}. This normalization
condition was used in 
our previous studies \cite{Bazavov:2011nk, HotQCD:2014kol, Bazavov:2018wmo}. The 
normalization constant depends on the amount of smearing, i.e. the coefficient $2c_Q$ of 
the $1/a$ divergence is smearing dependent. The larger the amount of smearing, the smaller 
the coefficient of the $1/a$ divergence becomes. For unsmeared Wilson line correlators the 
coefficient $c_Q$ was determined in Ref. \cite{Bazavov:2018wmo} for several beta values 
including, the two lowest ones used here, namely $c_Q(\beta=7.596)=0.3545(11)$ and 
$c_Q(\beta=7.825)=0.3403(12)$. Interpolating the results for $c_Q$ from Ref. \cite{Bazavov:2018wmo}
with cubic polynomial we estimate $c_Q(\beta=8.249)=0.3144(10)$. 

\begin{figure}
\centering
\includegraphics[width=0.45\textwidth]{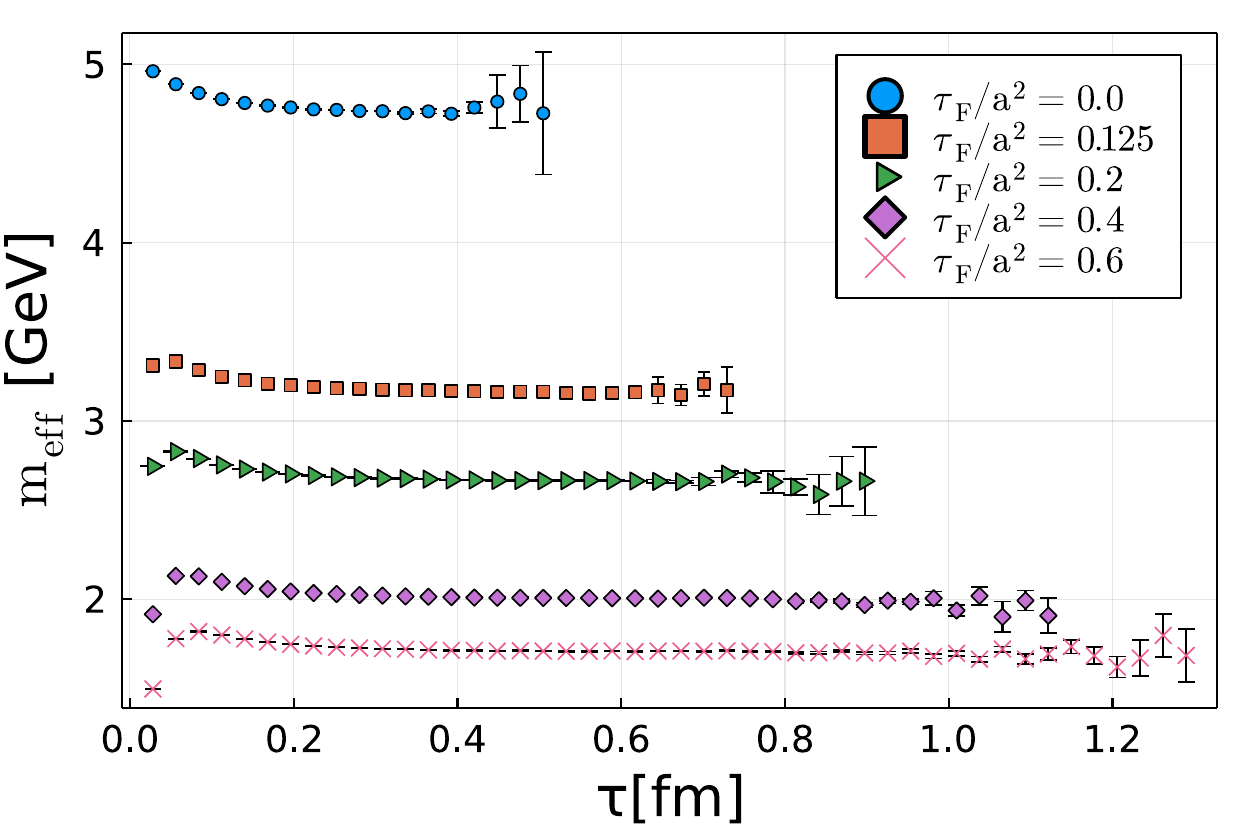}
\includegraphics[width=0.45\textwidth]{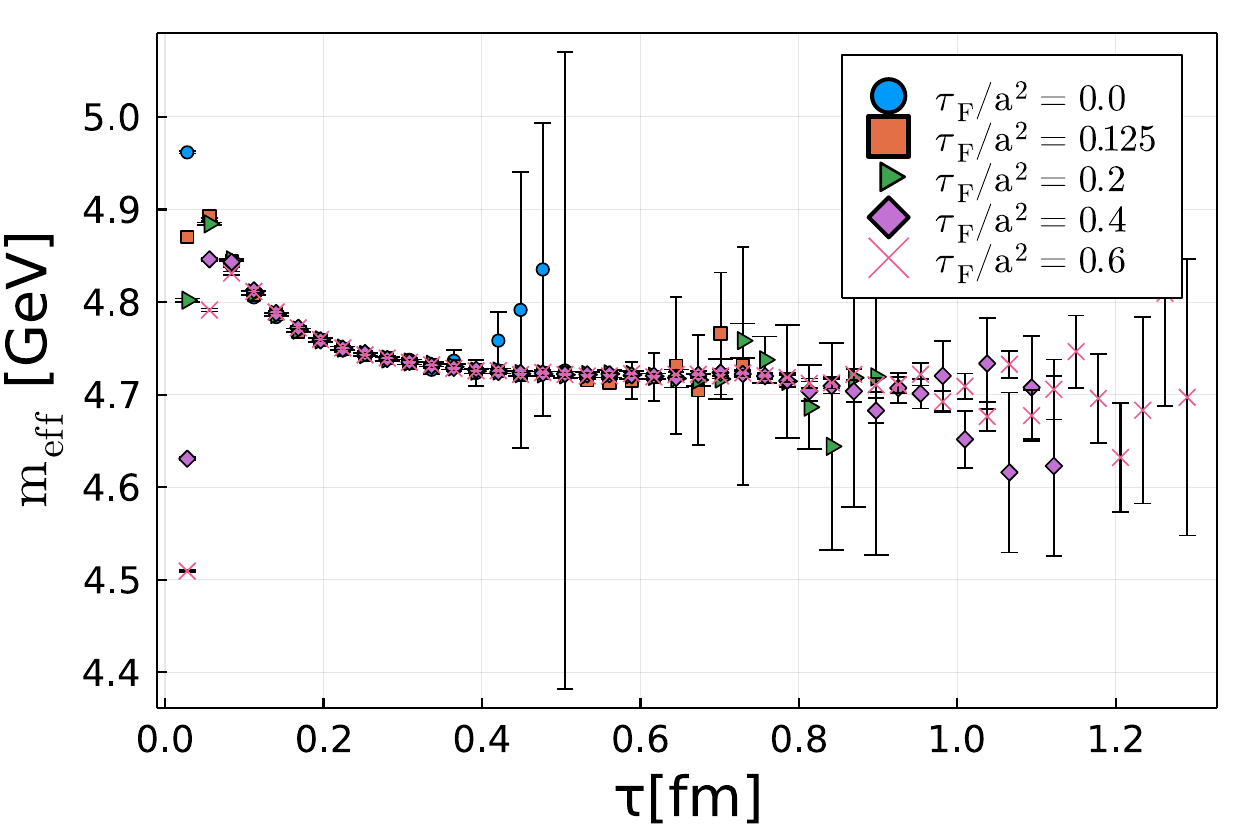}
\caption{The effective masses corresponding to the Wilson line correlators at $r/a=15$,
$\beta=8.249$ obtained for different flow times (top). The effective masses for different flow
times after applying the additive shift are discussed in the text (bottom).}
\label{fig:smear_comp}
\end{figure}

\begin{figure}
\centering
\includegraphics[width=0.45\textwidth]{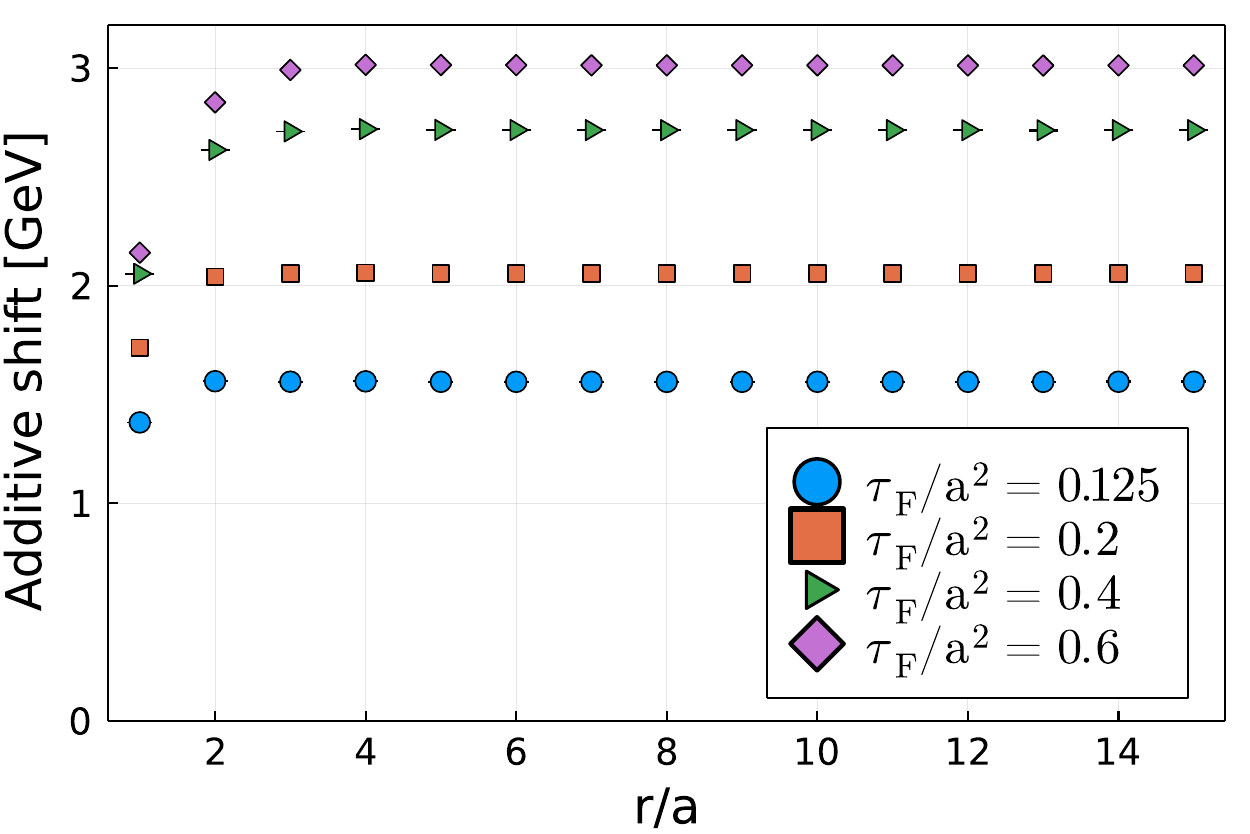}
\caption{The additive shifts for different flow times as a function of $r/a$ for $\beta=8.249$.}
\label{fig:shift}
\end{figure}

\begin{figure}
\includegraphics[width=0.45\textwidth]{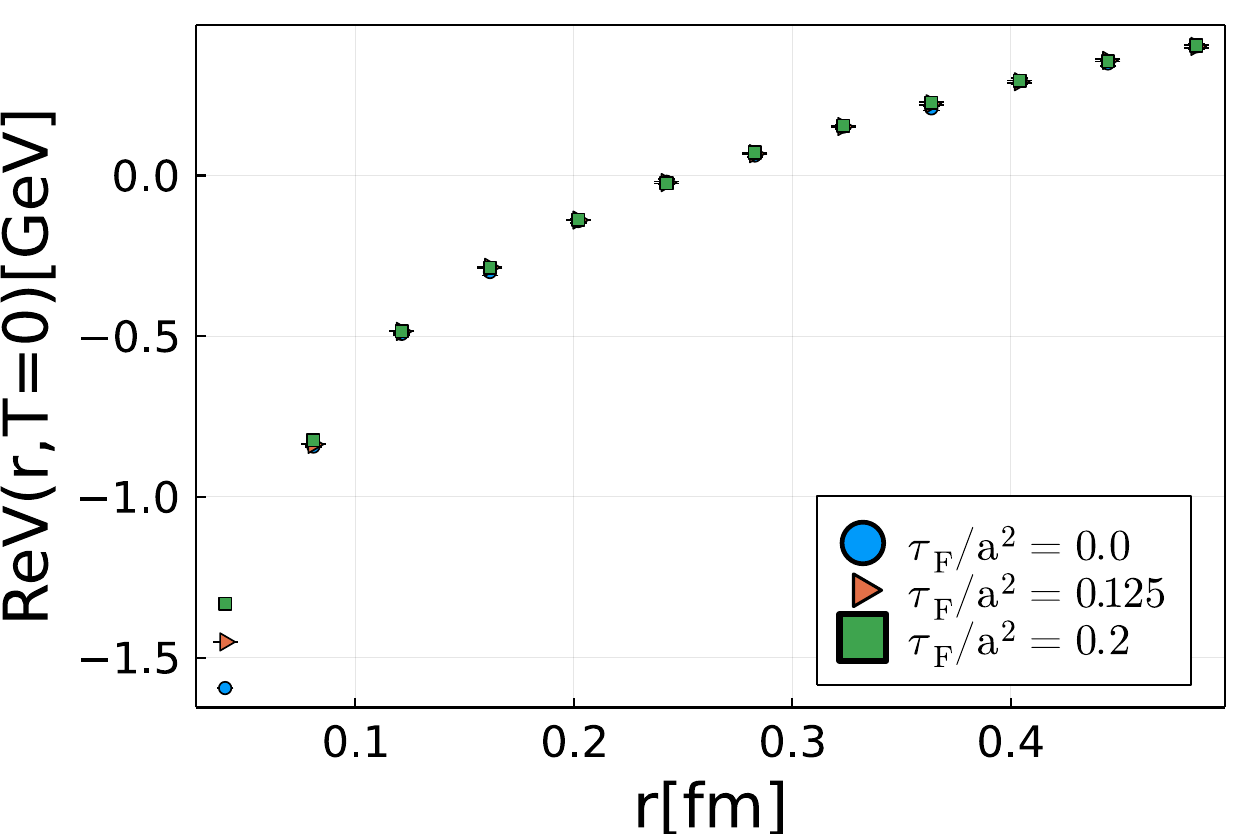}
\caption{The zero temperature potential for $\beta=8.249$ obtained with different smearing levels, including no smearing.}
\label{fig:pot_dist}
\end{figure}

In Fig. \ref{fig:smear_comp} (top) we show the un-renormalized effective masses at $T=0$
for $\beta=8.249$ at different flow times. 
The improvement in the signal 
with increasing flow time 
% signal 
at large $\tau$ is obvious from the figure.
We also see that the effective masses decrease with 
increasing flow time as one would expect based on the discussions above. There is a non-monotonic 
behavior of the effective masses in $\tau$ for $\tau/a=1-3$. This is due to the fact that the 
gradient flow distorts short distance physics and potentially can lead to non-positive definite
spectral function for very large $\omega$. However, for not too large $\omega$ there is no sign
of positivity violation in the spectral function since the effective masses approach plateaus from
above for $\tau/a>3$. This means that the gradient flow does not lead to artifacts in the 
determination of the $Q \bar Q$ potential at $T=0$. 
With a constant, flow-time dependent shift the  effective masses for different $\tau_F$ can be collapsed 
% By shifting the effective masses for different $\tau_F$ by a constant it is possible to collapse them 
to one line, except for very small $\tau$, where there are $\tau_F$-dependent distortions due to gradient flow. This is demonstrated in 
Fig. \ref{fig:smear_comp} (bottom). We determine this shift by fitting the difference in the 
effective masses calculated at different flow times to a constant for $\tau/a=7-18$ for 
$\beta = 8.249$ and $\tau/a=7-15$ for the two smaller values of $\beta$. This constant shift 
should amount to the difference in the additive normalization of the $Q\bar Q$ potential, and 
therefore, should be independent of $Q\bar Q$ separation, $r$, apart from the distortions at 
small $r$ due to smearing. In Fig. \ref{fig:shift} we show the relative shifts as a function of
$r$ for $\beta=8.249$ . We see that for very small $r$ there is some dependence on the value of $r$
implying that there are distortions in the zero temperature potential at these
distances due to smearing as expected. Namely, when $\tau_F/a^2 \le 0.2$ we see distortion
for $r/a < 2$, while for larger flow time we see distortions for $r/a <3$. To demonstrate 
this in Fig. \ref{fig:pot_dist} we show the zero temperature potential for different smearing levels
for relatively small $r$ values.
We see from the figure that that except for the smallest distance the potential does not
depend on the smearing level including the case of no smearing. 
We found that the situation for the other two $\beta$ values is the same.

\begin{figure}
\centering
\includegraphics[width=0.45\textwidth]{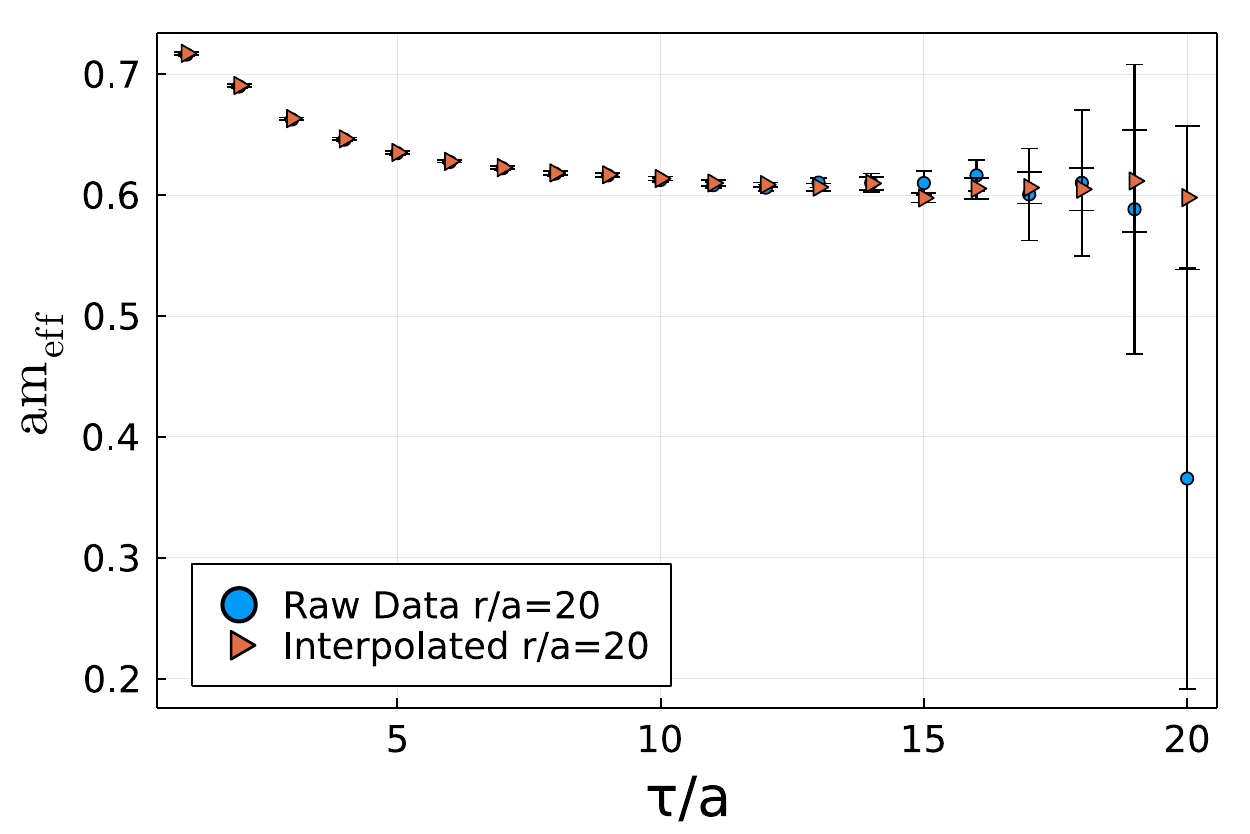}
\includegraphics[width=0.45\textwidth]{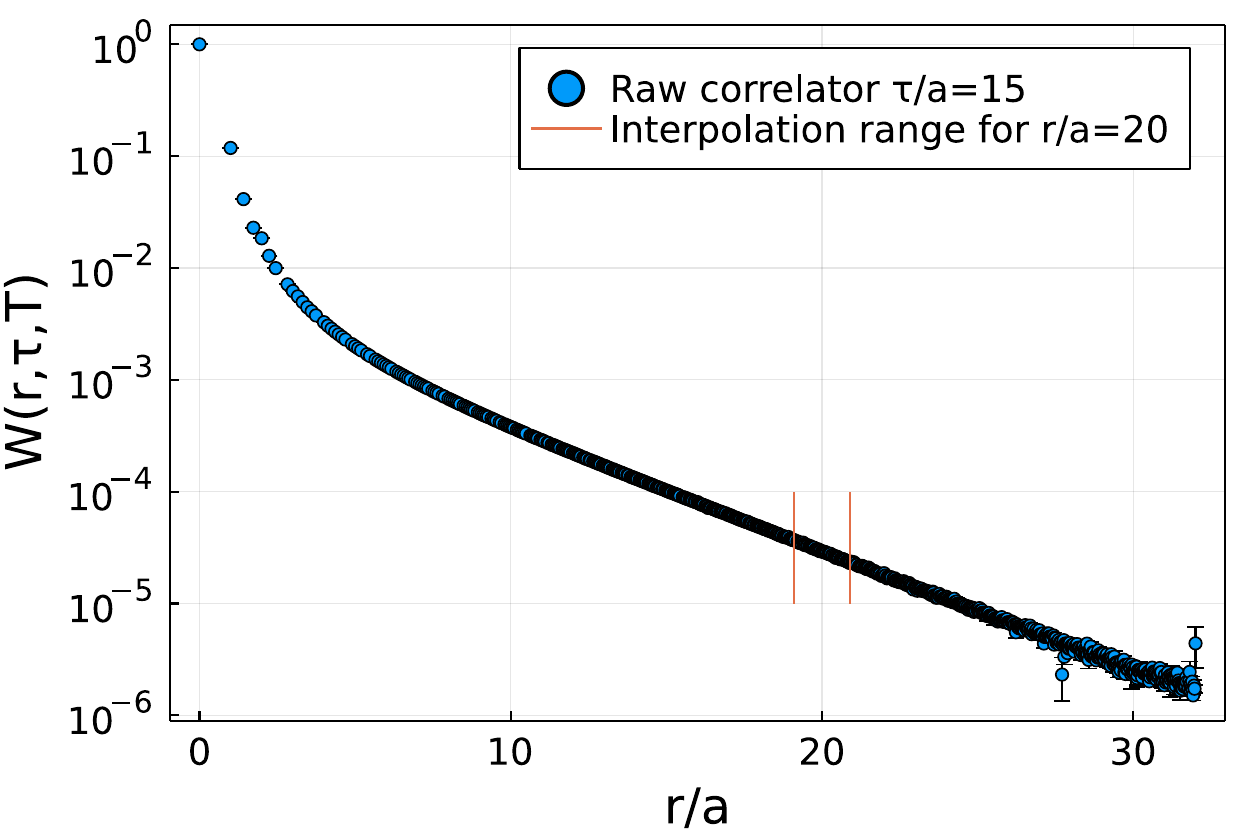}
\caption{(top) Effective mass for $N_\tau=64$, $N_x=64$, $r/a=20$ and $\tau_F/a^2=0.125$ for the raw data, 
compared to an interpolation fit done around $r/a=20$ in a range 
$\pm \Delta r/a=0.9$ with a second order polynomial. (bottom) The correlator as a function of distance r at fixed $\tau=15a$ for the same lattice as the top plot. }
\label{fig:inter}
\end{figure}

In addition to the gradient flow, we use polynomial interpolations to reduce fluctuations in
the Wilson line correlators. For fixed $\tau$ the Wilson line correlators should be a smooth
function of $r$ apart from the effects of breaking of rotational symmetry on the lattice.
For Symanzik gauge action these effects are smaller than the statistical errors for $r/a>3$
\cite{Bazavov:2014soa,Bazavov:2019qoo}. Therefore, it is natural to require that the data on
the Wilson line correlators are smooth functions of $r$ at a fixed value of $\tau$. By imposing
this requirement we effectively reduce the fluctuations in the original data set since nearby 
$r$ values usually correspond to very different path geometries and thereby suffer from quite 
independent gauge noise.
We perform second order polynomial  interpolations in a limited range of distances, $\Delta r$ 
around a target value of $r$ and replace the original datum with the interpolated value. 
We take into account that, with increasing distance, there are many different separations that 
are close to the target value of $r$ and adjust $\Delta r$ as we vary $r$. This additional 
noise reduction and the interpolation procedure are demonstrated in Fig. \ref{fig:inter}. In fact 
the result on the effective masses shown in Fig. \ref{fig:smear_comp} also incorporate the noise 
reduction from the interpolations. 
Because of the use of the above noise reduction 
the determination of the $Q\bar Q$ potential at 
zero temperature is now more accurate. Therefore,
we re-calibrated the central value of the constant $c_Q$ and used the
following values in the present analysis:
$c_Q(\beta=7.596)=0.3552$, $c_Q(\beta=7.825)=0.3401$
and $c_Q(\beta=8.249)=0.3135$. These values agree with the one quoted above within errors.

\begin{figure}[H]
\centering
\includegraphics[width=0.45\textwidth]{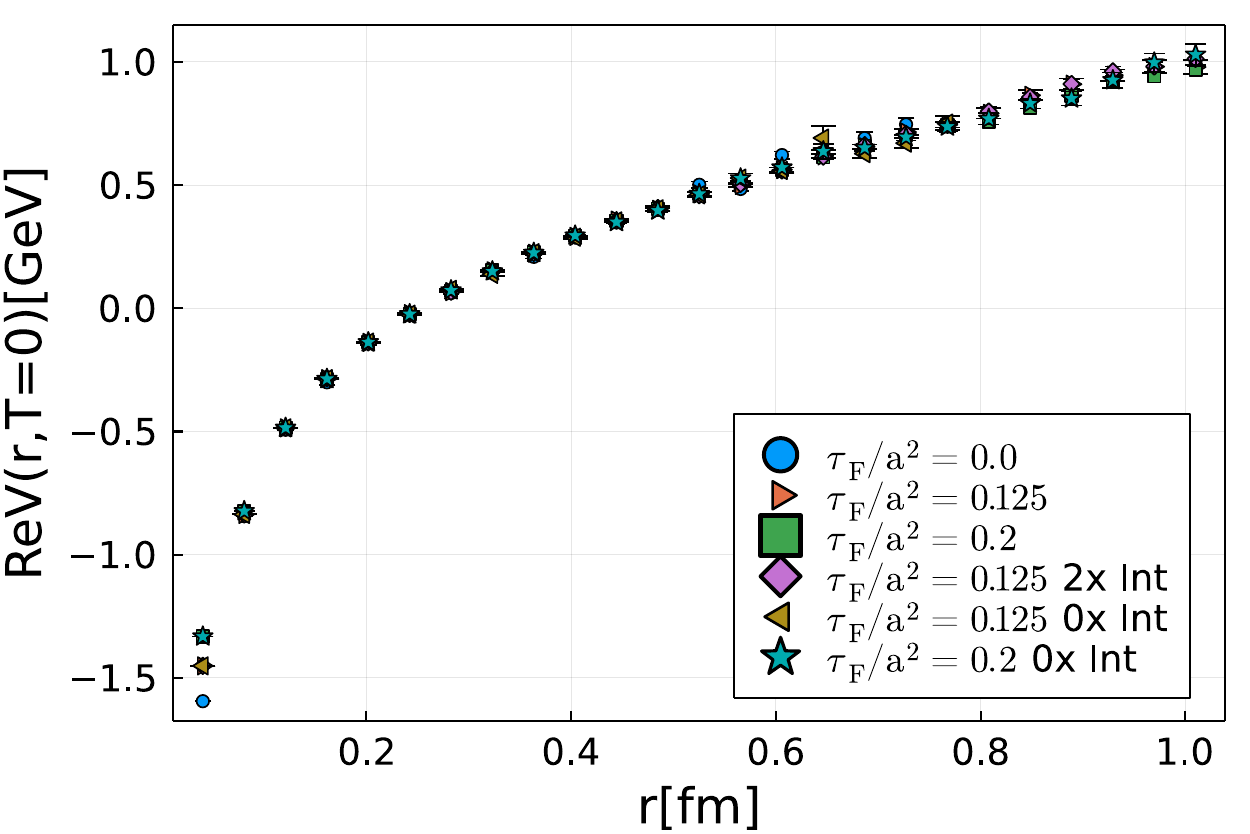}
\caption{
The zero temperature potential for different smearing and interpolation 
levels for $a=0.0404fm$. The label "$0 \times$ int" means no interpolation used in the analysis.
The label "$2 \times $ int" means that the $r$-interval used in the interpolation was doubled compared
to the default setup.}
\label{fig:comp0}
\end{figure}

To check that interpolations do not introduced additional bias we  performed
the analysis by doubling the interpolation range in $r$, and
also obtained the zero temperature potential without any interpolations. The results are shown in Fig. \ref{fig:comp0}.
As one can see the zero temperature potential is not sensitive to these changes. Doubling the interval in the interpolations
does not change the result, while skipping the interpolation in the analysis only results in large statistical
fluctuations.

\section{Analysis of the Wilson line correlators at $T>0$}
\label{app:analT}

\begin{figure}[H]
\centering
\includegraphics[width=0.45\textwidth]{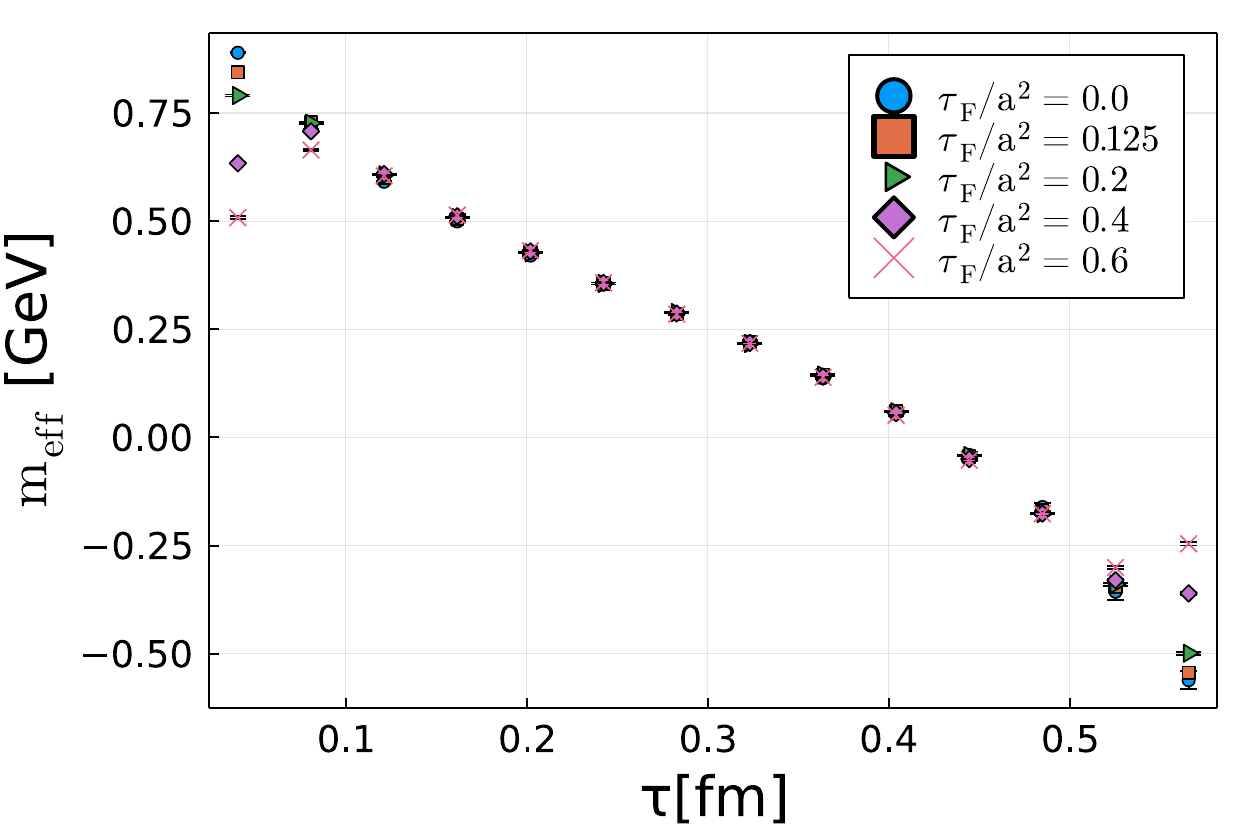}
\includegraphics[width=0.45\textwidth]{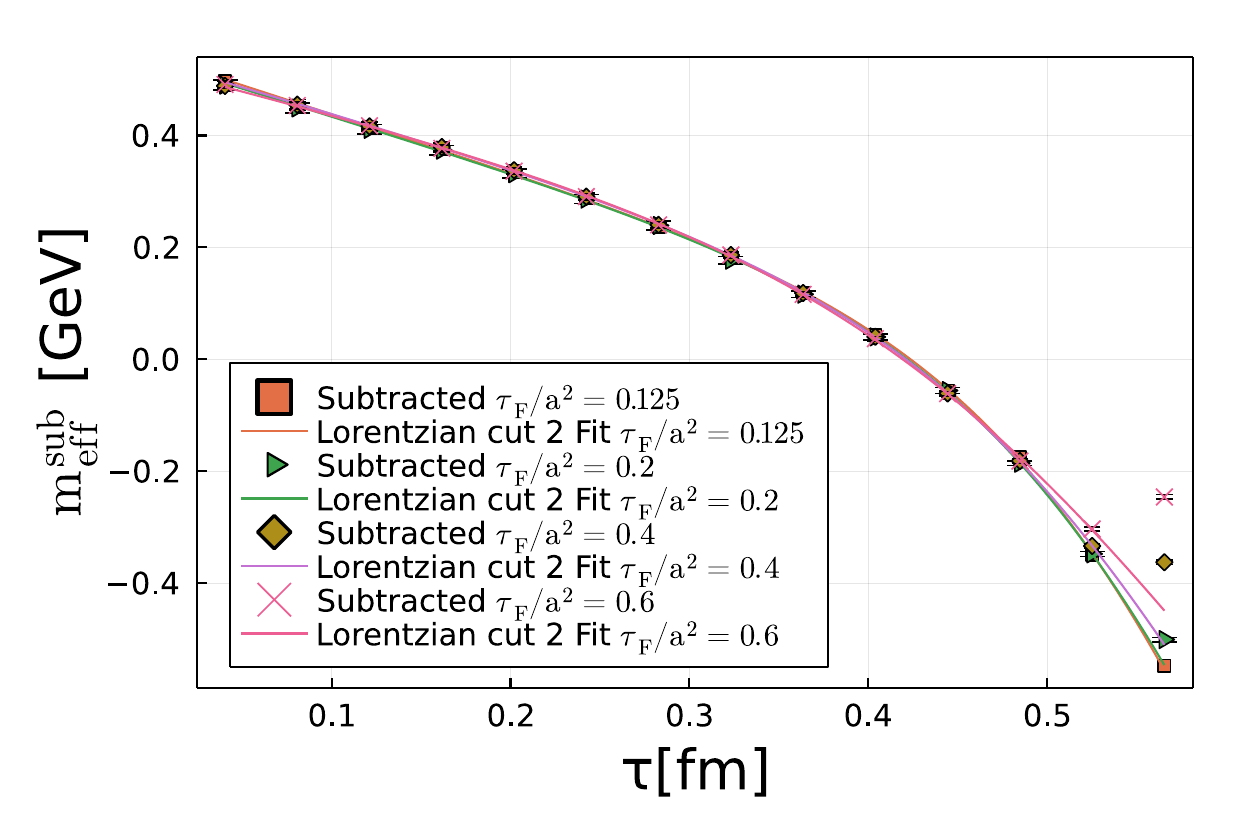}
\caption{The effective masses
for different flow times at $T=305$ MeV, $r=0.606$ fm, $\beta = 7.825$.
The bottom panel shows the effective masses for the subtracted
correlator. The lines in the bottom panel show the fits
discussed in the text.
}
\label{fig:smear_comp2}
\end{figure}

In this appendix we discuss the analysis of the Wilson line correlators at $T>0$
Our aim is to gain information on the spectral function corresponding to the Wilson line correlator 
at $T>0$. 
As discussed in the main text we use  the following Ansatz for the spectral function 
 \begin{equation}
     \rho_r(\omega,T)=
     \rho_r^{\text{low}}(\omega,T)+\rho_r^{\text{peak}}(\omega,T)+\rho_r^{\text{high}}(\omega),
 \end{equation}
 where $\rho_r^{\text{high}}(\omega)$ is the dominant part of the spectral function at large 
 $\omega$ and is assumed to be temperature independent. Furthermore, $\rho_r^{\text{peak}}(\omega,T)$ 
 describes the dominant peak in the spectral function and encodes the complex potential at $T>0$, 
 while $\rho_r^{\text{low}}$ is a small contribution to the spectral function below the dominant 
 peak, which is discussed below in more detail. 
The position and width of the dominant peak in 
the spectral function should not depend on the interpolating operator details used in the static 
$Q \bar Q$ correlator, e.g. on the flow time and whether we use Wilson line correlators in Coulomb 
gauge or Wilson loops. On the other hand $\rho_r^{\text{low}}(\omega,T)$ and 
$\rho_r^{\text{high}}(\omega)$ will depend on the specific choices of the interpolating operators 
used in the correlator, e.g. on the amount of smearing or the gauge tolerance used. In 
Fig. \ref{fig:smear_comp2} we show the effective masses for $T=305$ MeV and $r=0.606$ fm for 
different flow times. We see non-monotonic behavior and flow time dependence for small $\tau$ as 
we do for the $T=0$ case. However, for an intermediate $\tau$-range $0.1~{\rm fm}<\tau<0.45$ fm, where 
the contribution from $\rho_r^{\text{peak}}(\omega,T)$ is the dominant one, the effective masses for 
different flow times agree with each other very well. At $\tau>0.5$ fm the contribution from 
$\rho_r^{\text{low}}(\omega,T)$ becomes important, and we see some dependence on the flow time.
As discussed in Ref. \cite{Bala:2021fkm} $\rho_r^{\text{low}}(\omega,T)$ depends on the overlap
of the chosen $Q\bar Q$ operator with the light states that propagate backward in the Euclidean
time together with the forward propagating $Q\bar Q$. Similar dependence on the level of spatial
link smearing of the effective mass was observed in Ref. \cite{Bala:2021fkm}.

As discussed in the main
text the effective masses corresponding to
$W^{\text{sub}}(\tau,r,T)$ decrease monotonically with $\tau$, and
for sufficiently small $\tau$ they are approximately linear 
in $\tau$. This is demonstrated in Fig. \ref{fig:smear_comp2},
where the effective masses from $W^{\text{sub}}(\tau,r,T)$ are shown
for $T=305$ MeV and $r=0.606$ fm at various flow times. Thus the removal of the high
energy part of the spectral function also removes the artifacts 
induced by the gradient flow. 
For a small contribution from $\rho_r^{\text{low}}(\omega,T)$ this
%The 
linear behavior of the effective
masses in $\tau$ for small $\tau$ could
%can 
be easily explained if $\rho_r^{\text{peak}}(\omega,T)$ had
%has 
a Gaussian form $\rho_r^{\text{G}}(\omega,T) \sim e^{-(\omega-V(r,T))^2/(2\Gamma_G ^2)}$
% and the contribution from $\rho_r^{\text{low}}(\omega,T))$ is small 

\begin{align}
   W^{\text{sub}}(\tau,r,T) &\sim 
   \int d \omega e^{-\omega \tau }\rho_r^{\text{G}}(\omega,T)
   \nonumber\\
    &\sim \exp(-V(r,T)\tau+\frac{\Gamma_G^2}{2}\tau ^2).
\end{align}
However, the Gaussian form of the spectral function is not physically
motivated and the width of the Gaussian cannot be 
interpreted as ${\rm Im} V(r,T)$. If we assume that the detailed
shape of the spectral function away from the peak position is not
too important we can define 
%the 
${\rm Im} V(r,T)$ as the width
at half maximum height. In this case, a Gaussian form of the spectral
function can be used. A physically appealing choice of 
$\rho_r^{\text{peak}}(\omega,T)$ is a Lorentzian form. However, this
form is only valid for $\omega$ values that are not too far
from $\omega={\rm Re} V(r,T)$. The HTL spectral function of 
static $Q\bar Q$ \cite{Burnier:2013fca} is Lorentzian
only in the vicinity of the peak  and decays exponentially
when $|{\rm Re} V-\omega|$ is larger \cite{Burnier:2013fca}.
The same holds for the spectral function in the T-matrix approach
\cite{Liu:2017qah}.  Therefore, we use a cut Lorentzian
for $\rho_r^{\text{peak}}(\omega,T)$ in our analysis
\begin{equation}
    \rho_r^{\text{cL}}(\omega) =
    \frac{1}{\pi}\frac{A_r\Gamma_L \theta(\text{Cut}-|\omega-{\rm Re} V|)}{(\omega - {\rm Re} V)^2+\Gamma _L ^2}.
\end{equation}
It turns out that the cut Lorentzian also gives an almost linear dependence in $\tau$ for the effective masses.
In our analysis,
we set $\text{Cut}=2 \Gamma _L$.
To cross-check our results we also use the Gaussian form. 

\begin{figure}
\centering
\includegraphics[width=0.45\textwidth]{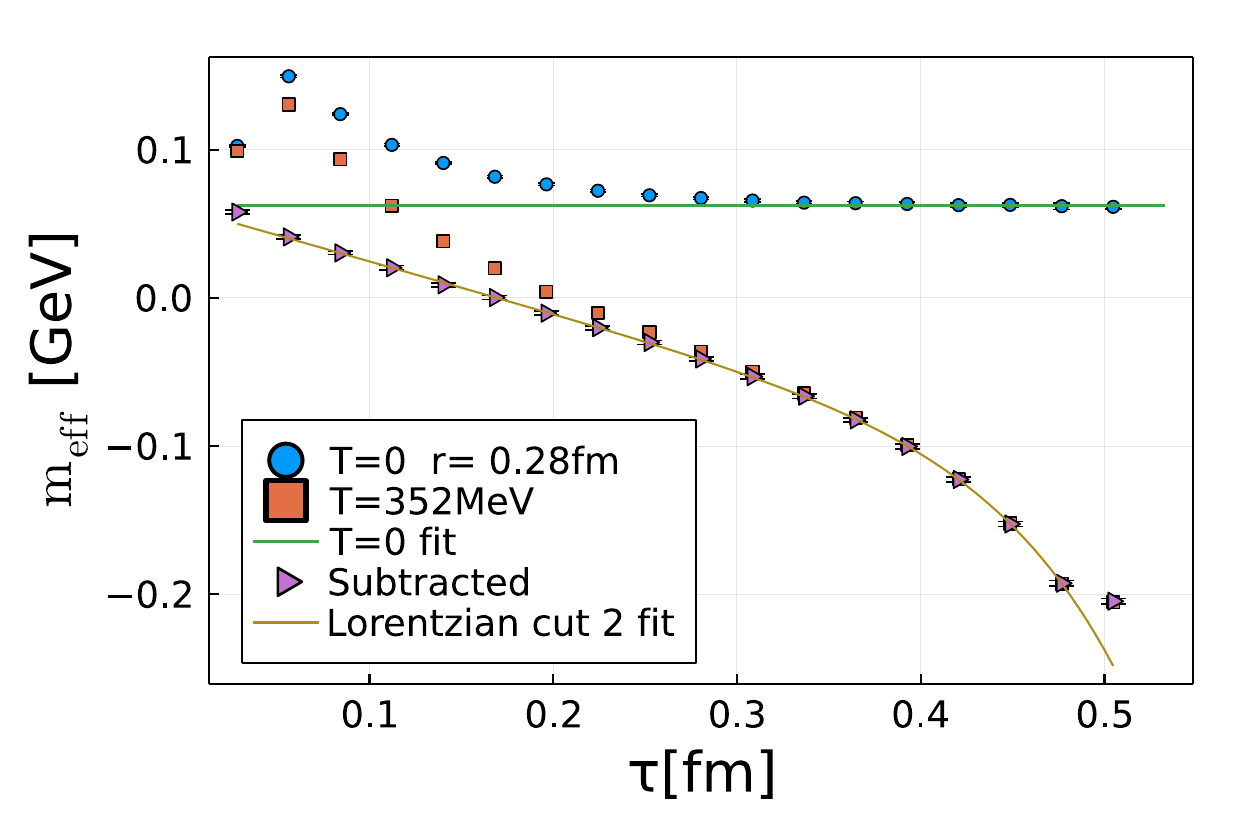}
\includegraphics[width=0.45\textwidth]{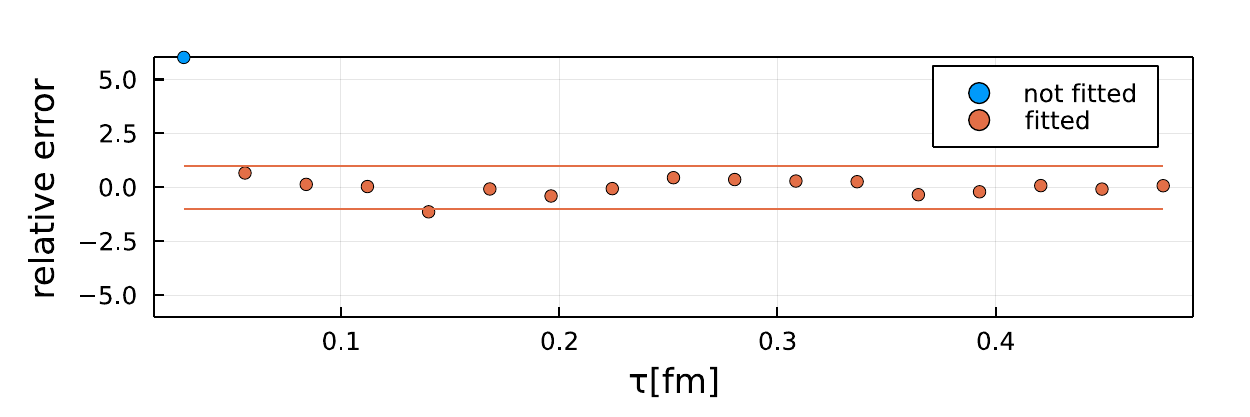}
\caption{The effective masses for $\beta=8.249$, $T=352$ MeV, $r=0.280$ fm
and the corresponding fits with the cut Lorentzian plus the delta function for $\rho_r^{\text{low}}$
shown as a line. The bottom panel shows the relative deviation between the fit and the data
with the lines indicating the estimated $1-\sigma$ band of the data.}
\label{fig:Nt20_sub}
\end{figure}
\begin{figure}
\centering
\includegraphics[width=0.45\textwidth]{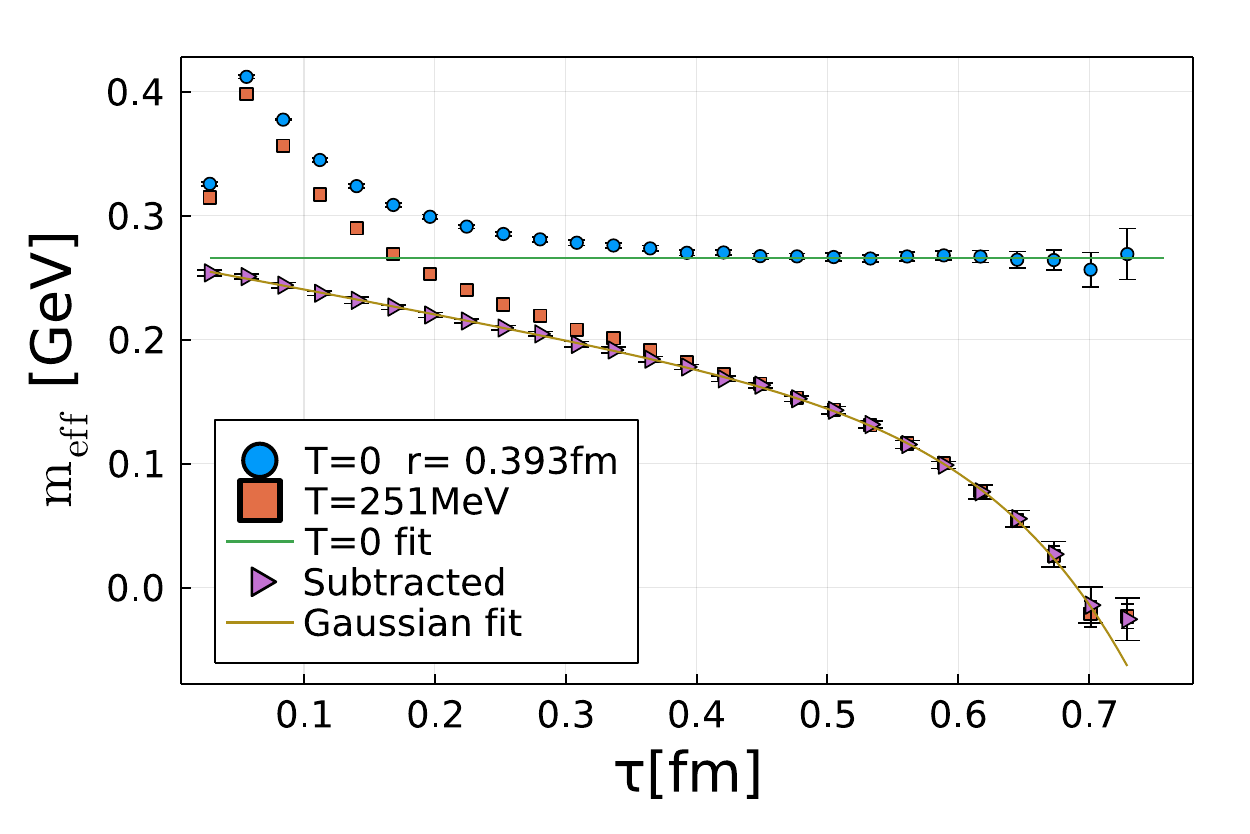}
\includegraphics[width=0.45\textwidth]{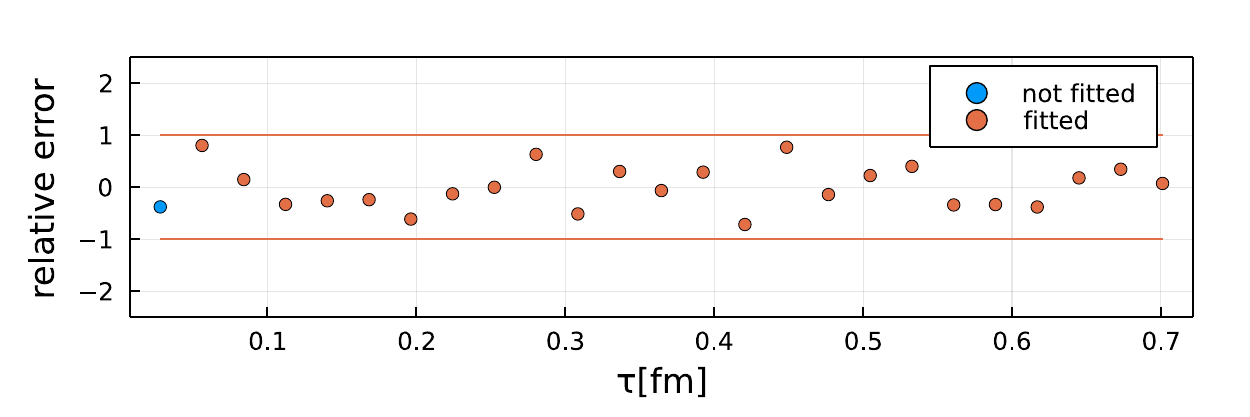}
%\caption{gaussian plus delta. $T=251 MeV$ $r/a=14$. }
\caption{The effective masses for $\beta=8.249$, $T=251$ MeV, $r=0.392$ fm
and the corresponding fits with a Gaussian plus the delta function for $\rho_r^{\text{low}}$
shown as a line. The bottom panel shows the relative deviation between the fit and the data
with the lines indicating the estimated $1-\sigma$ band of the data.}
\label{fig:Nt28_sub}
\end{figure}

It was shown in Ref. \cite{Bala:2021fkm} that the rapid non-linear decrease in the effective masses is due to $\rho_r^{\text{low}}(\omega,T)$.
This contribution to the spectral function arises from the light
states in the medium propagating backward in time which are coupled
to the static $Q\bar Q$ propagating forward in time
\cite{Bala:2021fkm}. This contribution also depends on the details 
of the $Q\bar Q$ correlators, e.g. whether one uses Wilson line or
Wilson loops and the amount of smearing used \cite{Bala:2021fkm}.
We model this part of the spectral function with a single delta 
function because such a simple form is sufficient to describe
the data for the Wilson line correlators with the exception of one 
data point very close to the boundary $\tau=1/T$. We perform
fits of subtracted Wilson line correlator
with cut Lorentzian form of $\rho_r^{\text{peak}}(\omega,T)$ and
a single delta function for $\rho_r^{\text{low}}(\omega,T)$ for
all available data sets omitting the first datum, which is possibly
affected by the distortions due to smearing, and the last data point.
Some sample fits  are shown in Fig. \ref{fig:smear_comp2} for $T=305$ 
MeV, $r=0.606$ fm, $\beta = 7.825$, and
in Fig. \ref{fig:Nt20_sub}  for $T=352$ MeV, $r=0.28$ fm, $\beta=8.249$. The fits work well as demonstrated
in Fig. \ref{fig:Nt20_sub} (bottom), where the relative difference between the fit and the lattice
data is shown. Fits using the Gaussian form for $\rho_r^{\text{peak}}(\omega,T)$ work equally well as
demonstrated in Fig. \ref{fig:Nt28_sub}. 

\begin{figure*}
\includegraphics[width=0.45\textwidth]{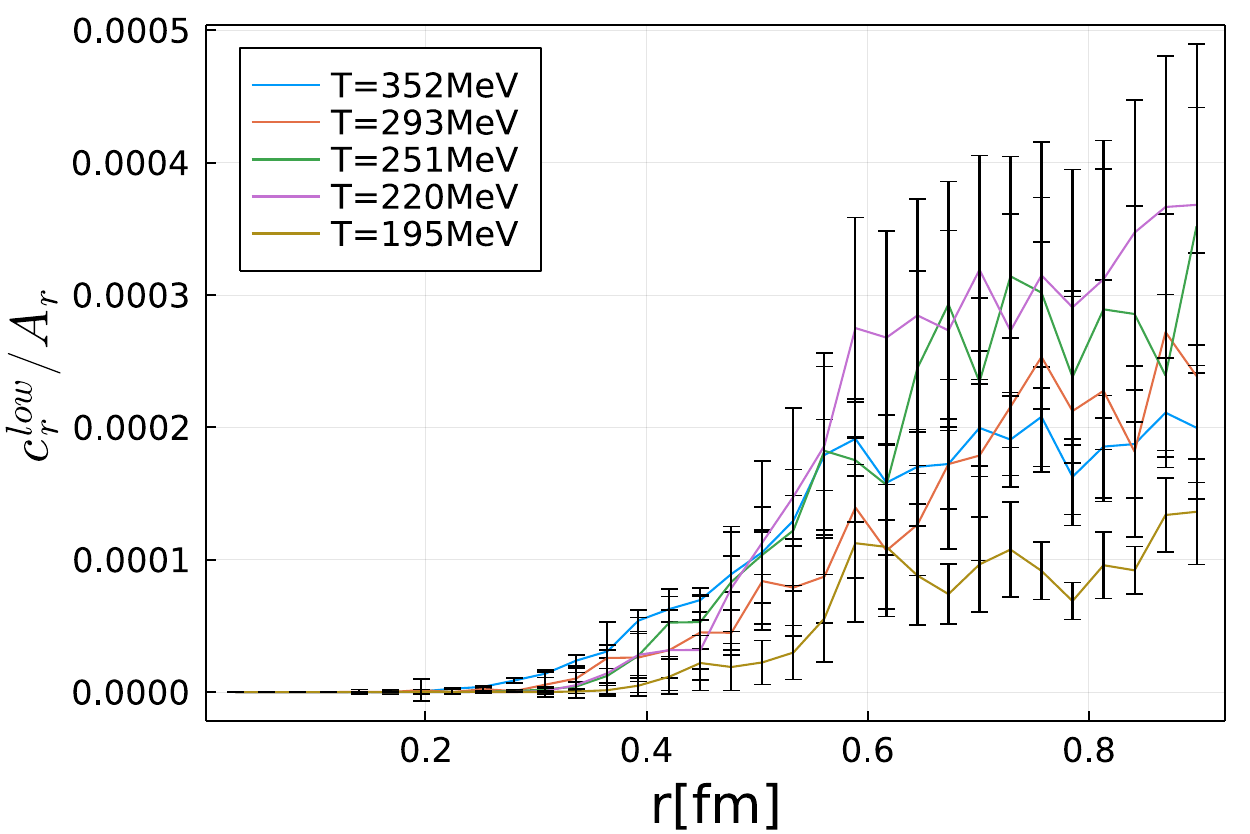}
\includegraphics[width=0.45\textwidth]{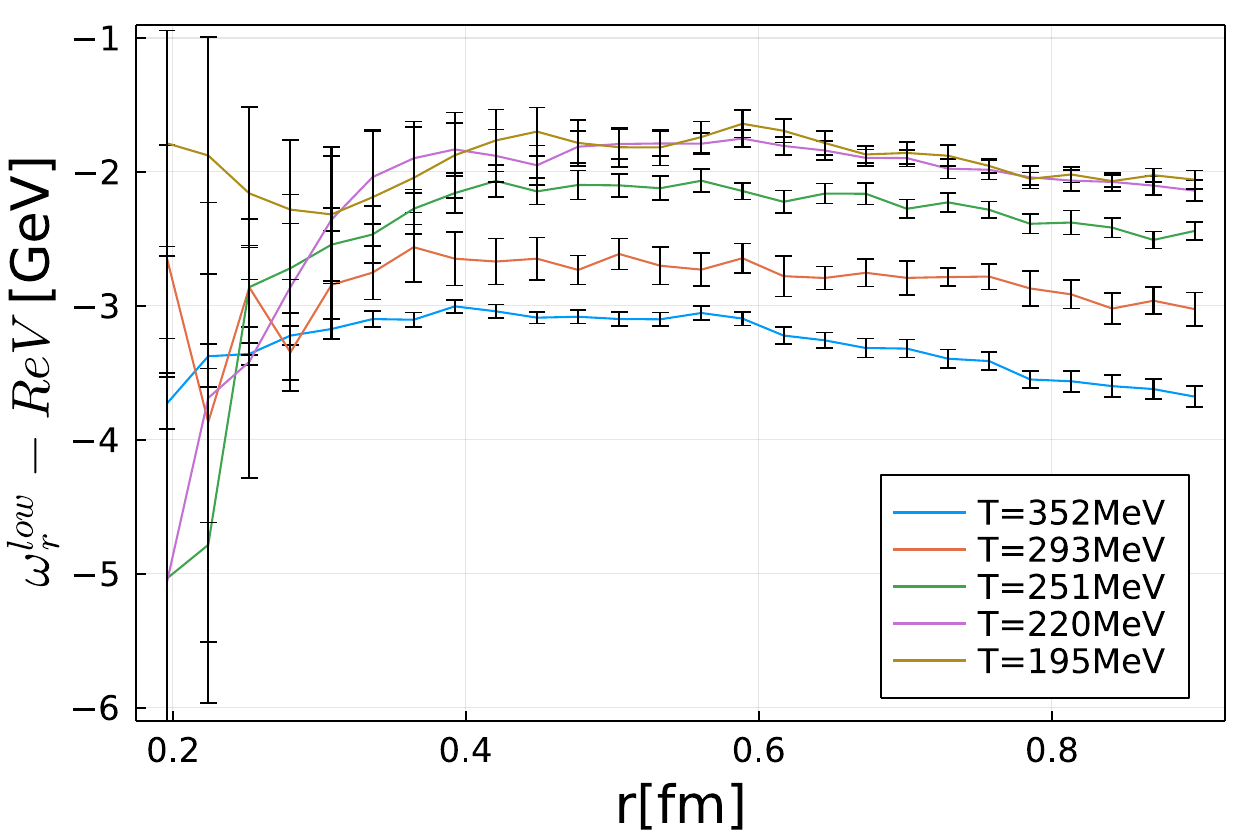}
\caption{The amplitude of the small delta function divided by $A_r$ (left) and the position of the small delta function relative to the position of
the dominant peak (right) as a function of $r$. The results are shown at different temperatures
for lattice spacing $a=0.0280$ fm ($\beta=8.249$).}
\label{fig:delta_func}
\end{figure*}

The amplitude and the position of the small delta function that parametrizes $\rho_r^{\rm low}$ are shown relative to the dominant peak in Fig. \ref{fig:delta_func}
for $\beta=8.249$ and different temperatures. 
As one can see from the figure, the position of this delta function is between
$1.8$ GeV and 3.8 GeV  below the position of the dominant peak, and shows only mild dependence on $r$. The amplitude of this
delta function on the other
hand increases rapidly with increasing $r$. Similar results have been obtained for
the two other $\beta$ values.
We also note that for small values of $r$, typically smaller than five times the lattice spacing, it is not necessary to include this small delta function in the fits, 
i.e. we can set $\rho_r^{\rm low}$ to zero and obtain good fits.

\begin{figure}
\centering
\includegraphics[width=0.45\textwidth]{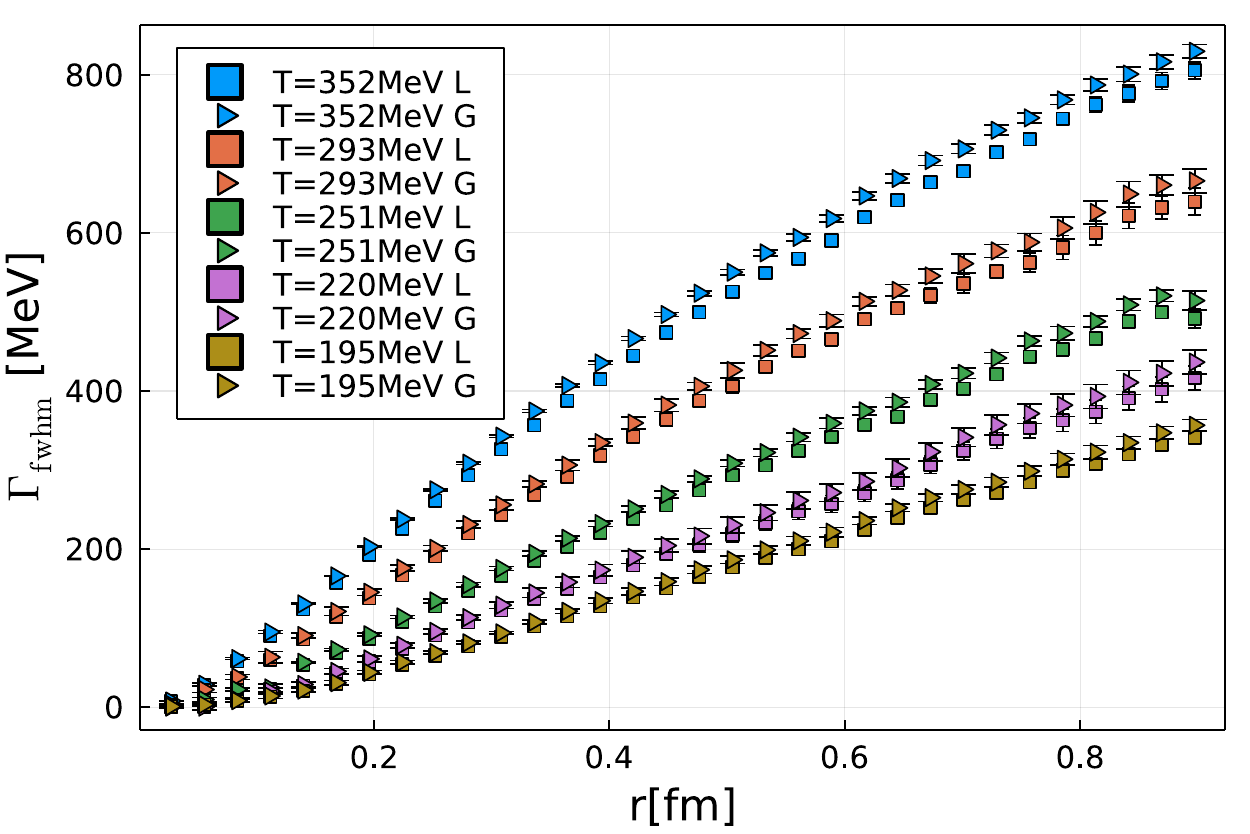}
\caption{Width at half the maximum height (FWHM) for the cut Lorentzian fit and Gaussian fit.}
\label{fig:Width_L_vs_G}
\end{figure}

In Fig. \ref{fig:Width_L_vs_G}
we show the width of the spectral function defined as the width at half of
the maximum height as a function of $r$ and different temperatures obtained
from the fits using Gaussian and cut Lorentzian form for $\rho_r^{\text{peak}}(\omega,T)$. We see that using the Gaussian results in a systematically larger
width. The Lorentzian parameter $\Gamma _L$ though is dependent  on the cut on the Lorentzian. This means that there is a systematic uncertainty in the determination
of ${\rm Im} V(r,T)$ from the parametrization of the spectral function. As
we discuss in the section below it is possible to define the width in a model independent way by considering cumulants of the spectral function. 

\begin{figure}
\centering
\includegraphics[width=0.45\textwidth]{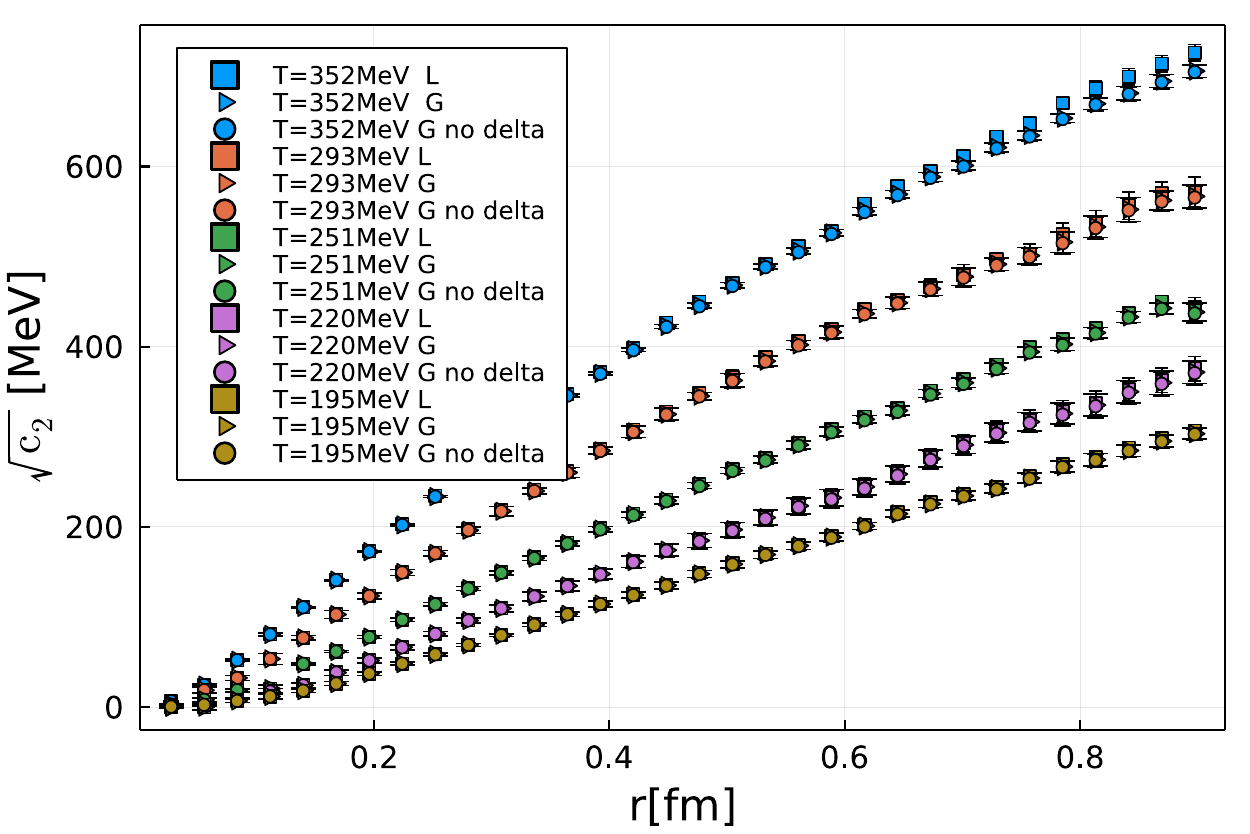}
\caption{$\sqrt{c_2}$ for the Gaussian fit (G), compared to the cut Lorentzian fit (L), or to a Gaussian fit without accounting for the low $\omega$ structure (G no delta).}
\label{fig:gaus_vs_lorentz_plus_width}
\end{figure}

\begin{figure}
\centering
\includegraphics[width=0.45\textwidth]{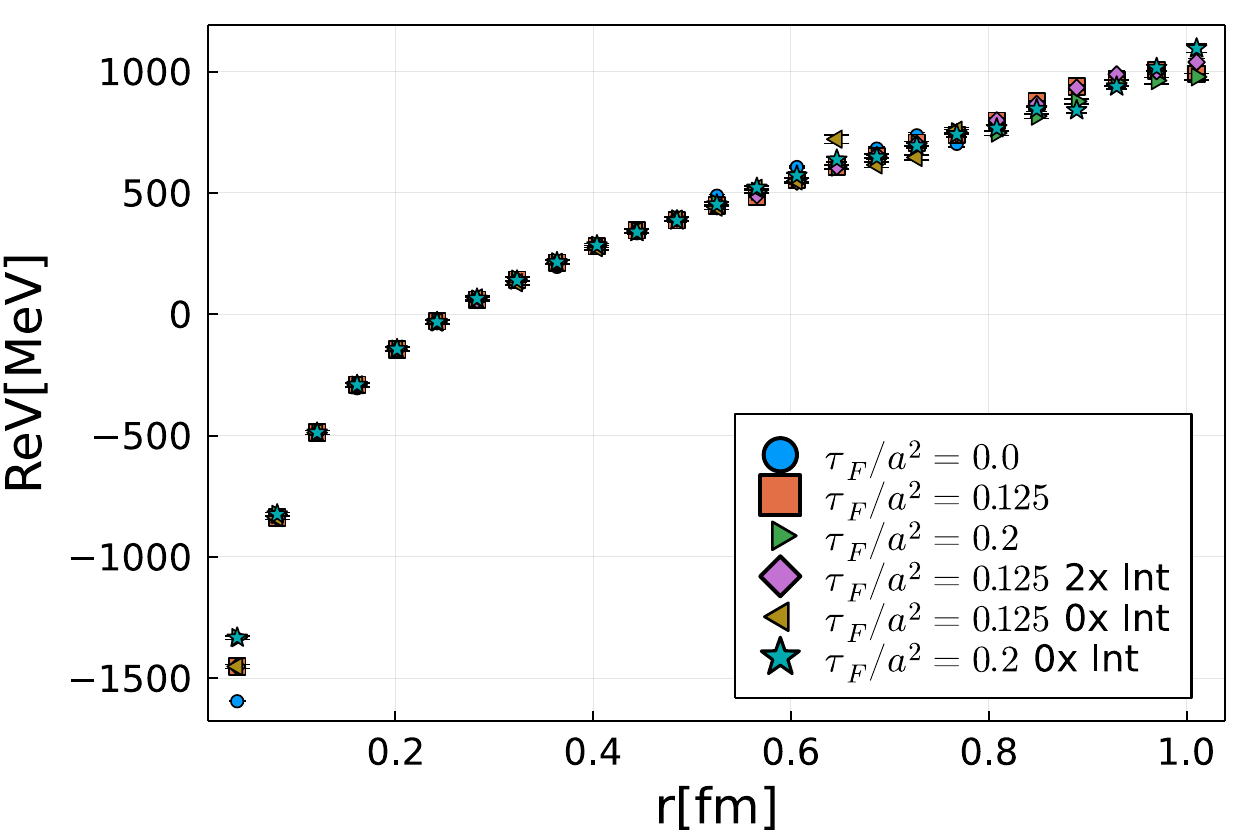}
\caption{Real part of the potential for the cut Lorentzian fit for different smearing and interpolation 
levels for $a=0.0404fm$, $N_\tau = 20$, $T=244$ MeV. The label "$0 \times$ int" means no interpolation
is used in the analysis, while the label "$2 \times$ int" means that the interpolations range was doubled in
the analysis compared to the default setup.
Fits done from $\tau=2$ to $17$. Fits with no smearing or no 
interpolation fits have been cut above $r/a\geq 20$ due to large errors. 
The fit for no smearing is only done up to $\tau=16$. Error bars are purely statistical. }
\label{fig:m1_comp}
\end{figure}

\begin{figure}
\centering
\includegraphics[width=0.45\textwidth]{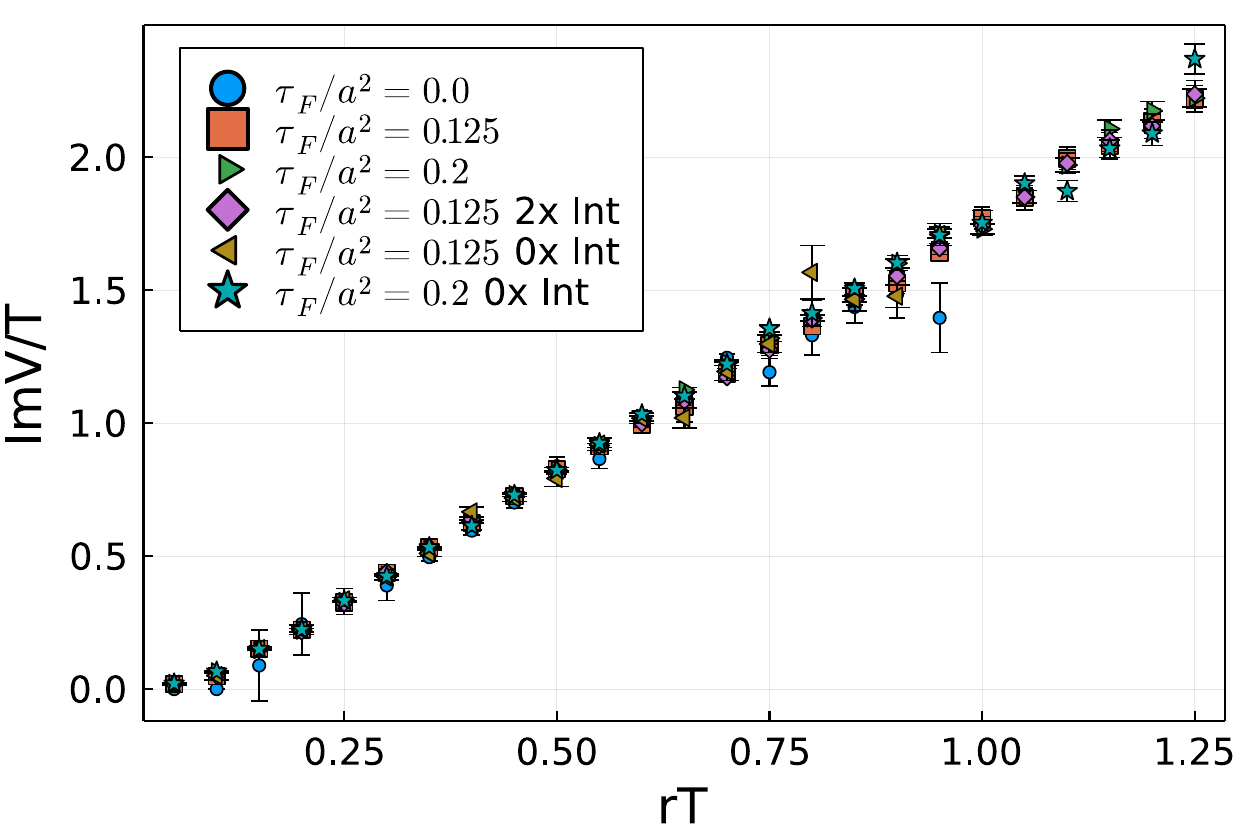}
\caption{Imaginary part of the potential for the cut Lorentzian fit for different smearing and interpolation 
levels for $a=0.0404fm$, $N_\tau = 20$, $T=244$ MeV. 
The label "$0 \times$ int" means no interpolation
is used in the analysis, while the label "$2 \times$ int" means that the interpolations range was doubled in
the analysis compared to the default setup.
Fits done from $\tau=2$ to $17$. Fits with no smearing or no 
interpolation  have been cut above $r/a\geq 20$ due to large errors.  
Fit for no smearing is only done up to $\tau=16$. Error bars are purely statistical.}
\label{fig:m2_comp}
\end{figure}

We also studied the dependence of our results on the real and imaginary part of the potential
on the number of smearing level and on the interpolations.
As in the zero temperature case we performed the analysis without using interpolation or
doubling the interpolation range. 
We  find that the distortions in ${\rm Re} V(r,T)$ due to smearing 
are the same as in the zero temperature potential. This is shown
in Fig. \ref{fig:m1_comp}. This means that  
while smearing can distort the potential at very short distances, it does not affect the temperature
dependence of the real part of the potential. Therefore, we show our results for ${\rm Re} V$ also at the shortest distances
in the main text. At these distance we use slightly different values of $c_Q$ shown in Fig. \ref{fig:shift} to
offset the distortions due to smearing. From Fig. \ref{fig:m1_comp} we also show that using interpolation
does not introduce a bias in our results for ${\rm Re} V(r,T)$.
The dependence of ${\rm Im} V(r,T)$ on the smearing level and on the interpolations is shown in Fig. \ref{fig:m2_comp}.
As one can see from the figure also here we do not see significant dependence on these.

\section{Cumulants of the spectral function}
\label{app:cum}

In this appendix we discuss the cumulants of the spectral
functions and their relation to the effective mass of the Wilson line correlators.
The cumulants of the spectral functions $c_n$ are defined as

\begin{align}
    c_1 &= \langle \omega\rangle , \\
    c_2 &= \langle \omega ^2\rangle -\langle \omega \rangle ^2, \\
    c_3 &= \langle \omega^3\rangle -3\langle \omega\rangle \langle \omega ^2\rangle + 2\langle \omega\rangle ^3, 
\end{align}
where $\langle ... \rangle$ stands for $\int d \omega \rho_r(\omega,T) ...$.
Cumulants exist if the spectral function has support in a finite $\omega$
range, which is the case for the subtracted spectral function 
$\rho_r^{\rm sub}(\omega,T)=\rho_r(\omega,T)-\rho_r^{\rm high}(\omega) \equiv \rho_r^{\text{low}}(\omega,T)+\rho_r^{\text{peak}}(\omega,T)$. 
In what follows we will discuss the moments of this spectral function.
The cumulants of the spectral function
are related to the cumulants of the subtracted Wilson line correlators at $\tau=0$, 
$m_n$ defined
as
\begin{equation}
    W^{\rm sub}(\tau,r,T)=\exp\Bigg[\sum_{n=0}^{\infty} \frac{m_n (-\tau)^n}{n!}\Bigg]  \label{eq:cumulants}
\end{equation}
%\begin{equation}
%    C(\tau) \approx C_{\text{ap}}(\tau) \JHW{\equiv} \exp(-a_1\tau+a2\tau^2)
%\end{equation}
%where ``{ap}'' stands for approximation.
%On the other hand, we have for small $\omega\tau$ that
This can be seen by Taylor expanding the exponential in
the spectral representation of the subtracted Wilson line correlator
\begin{align}
    W^{\rm sub}(\tau,r,T) 
    &= \int d \omega e^{-\omega \tau} \rho_r^{\text{sub}}(\omega,T) \nonumber\\
    &= \int d \omega 
    \sum_{n=0}^{\infty}\frac{(-\omega\tau)^n}{n!}
    % (1-\omega \tau+\frac{(\omega \tau)^2}{2}-\frac{(\omega \tau)^3}{6}+...)
    \rho_r^{\text{sub}}(\omega,T) \nonumber\\
    &=
    \sum_{n=0}^{\infty} \langle\omega^n\rangle \frac{(-\tau)^n}{n!}
    % 1-\langle \omega\rangle \tau+\langle \omega ^2\rangle \frac{\tau^2}{2}-\langle \omega ^3\rangle \frac{\tau^3}{6}+..., 
    \label{eq:expand}
\end{align}
%If we then expand $C_{ap}$ we get
%\begin{align}
%    C_{ap}(\tau) &= 1+(-a_1\tau+a2\tau^2)+\frac{(-a_1\tau+a2\tau^2)^2}{2} \nonumber\\
%    &+\frac{(-a_1\tau+a2\tau^2)^3}{6}+... \\
%    C_{ap}(\tau) &= 1-\tau a_1+\tau ^2 (\frac{a_1^2}{2}+a_2) + ..
%\end{align}
Expanding the exponential in Eq. \eqref{eq:cumulants} and comparing to Eq. \eqref{eq:expand} we see that:
%For the lowest two cumulants we have:
\begin{align}
    m_1 &= \langle \omega\rangle , \\
    m_2 &= \langle \omega ^2\rangle -\langle \omega \rangle ^2.
\end{align}
The first cumulant of the Wilson line correlators is the effective mass.
The second cumulant is the slope of the effective mass in $\tau$.

We calculated the second cumulant of the subtracted spectral function
using the Gaussian form and cut Lorentzian form including and excluding
the $\delta$ function at small $\omega$. The result of this analysis is shown
in Fig. \ref{fig:gaus_vs_lorentz_plus_width}. We see that the second cumulant of the 
spectral function is not sensitive whether we use a Gaussian or cut Lorentzian in our 
analysis. Furthermore, the second cumulant does not change
much if we include or exclude the contribution from $\rho_r^{\text{low}}(\omega,T)$ that is
the small delta peak. We also see that $\sqrt{c_2}$ has a similar dependence
on $r$ as the FWHM 
% width parameter shown 
in Fig. \ref{fig:Width_L_vs_G} but is
somewhat smaller. 

We have estimated the systematic uncertainties due to the Ansatz for the spectral function by varying the fit range for the 
cut Lorentzian Ansatz with $\sfrac{\tau_{\text{min}}}{a}=2$ and 
$\sfrac{\tau_{\text{max}}}{a}=\{N_\tau-5,N_\tau-4,N_\tau-3\}$. 
The effect of this variation for 
$\sqrt{c_2}$ exceeds the difference upon changing the Ansatz (to Gaussian) or the 
cut of the Lorentzian. Thus, we used 
the average of the two most outlying results for cut Lorentzian as central values 
and the full spread as the systematic error estimate, which we have added in quadrature. 
For large distances this estimate clearly exceeds the statistical errors.
${\rm Re} V(r,T)$ is insensitive to these changes within statistical errors.

\begin{figure}[H]
\includegraphics[width=0.45\textwidth]{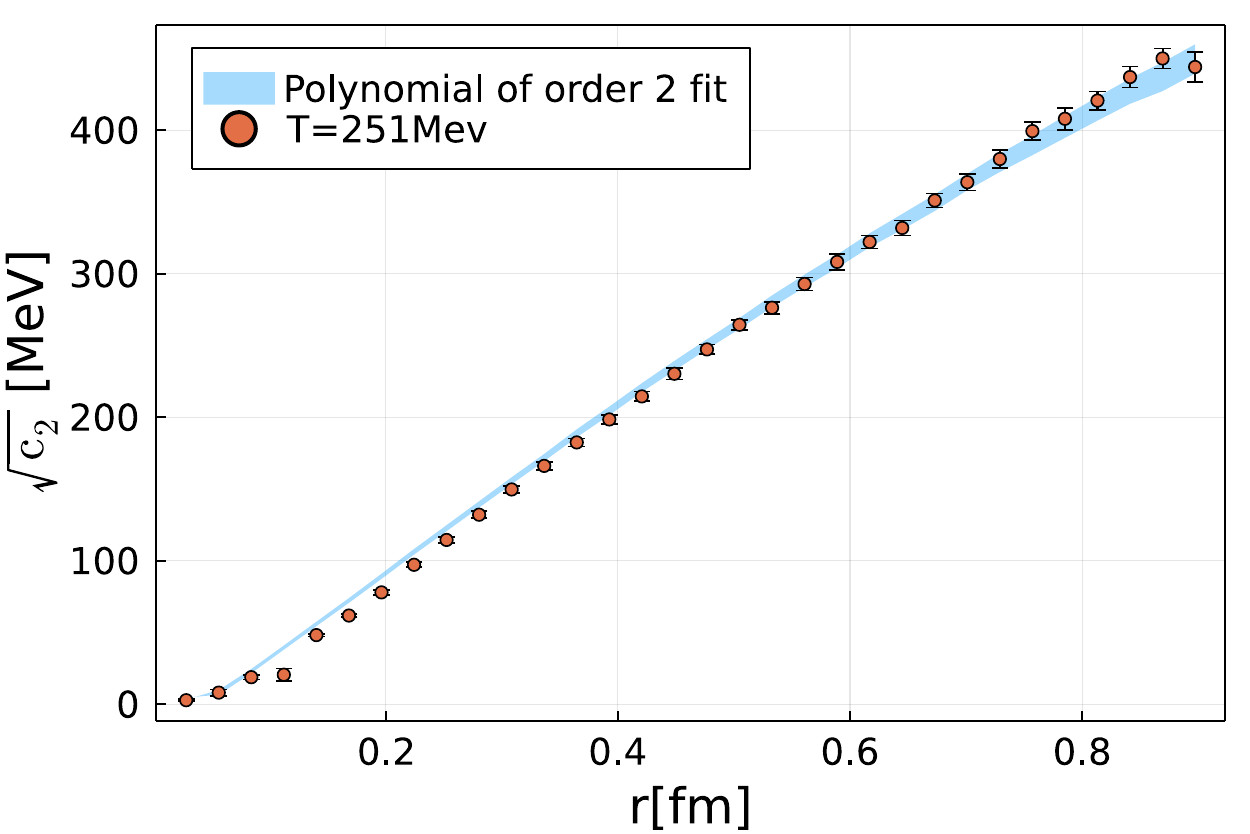}
\caption{The second cumulant of the subtracted
spectral function as a function of $r$ determined
from the cut Lorentzian form of the spectral
function (circles) and from the second order
polynomial fit of the Wilson line correlation
function in the $\tau /a$ range $2-N_\tau /3$ (blue
band) for $T=251$ MeV and $a=0.0280$ fm.
}
%Polynomial of order 2 fit in $\tau /a$ range $2-N_\tau /3$ to extract cumulants $c_i$, compared to the result for $c_2$ for the Lorentzian cut at 2 times the width. }
\label{fig:pol2}
\end{figure}

We can also fit our lattice results on the subtracted Wilson line correlator
with the following simple form
\begin{equation}
W^{\text{approx}}(\tau,r,T) = \exp(m_0-m_1\tau + m_2  \tau ^2/2)
\label{pol2}
\end{equation}
in the range 
$\tau /a=2-N_\tau /3$, where the effective mass is approximately linear. 
From this fit, we can then estimate the second cumulant of the spectral
function and compare it with the determination of $c_2$ obtained by
integrating the model spectral function based on the cut Lorentzian
and the small delta function in almost the entire $\tau$ range. This 
comparison is shown in Fig. \ref{fig:pol2}. We see that the two methods
of estimating $c_2$ are in good agreement. This means that defining
${\rm Im} V(r,T)$ in terms of $c_2$ is model independent and robust.
%We compare results between the polynomial extracted $c_2$ and that from 
%the cut Lorentzian fit for $T=251$ MeV in figure \ref{fig:pol2}, and find 
%that this totally model-agnostic fit produces almost the same outcome.}

\begin{figure}[H]
\includegraphics[width=0.45\textwidth]{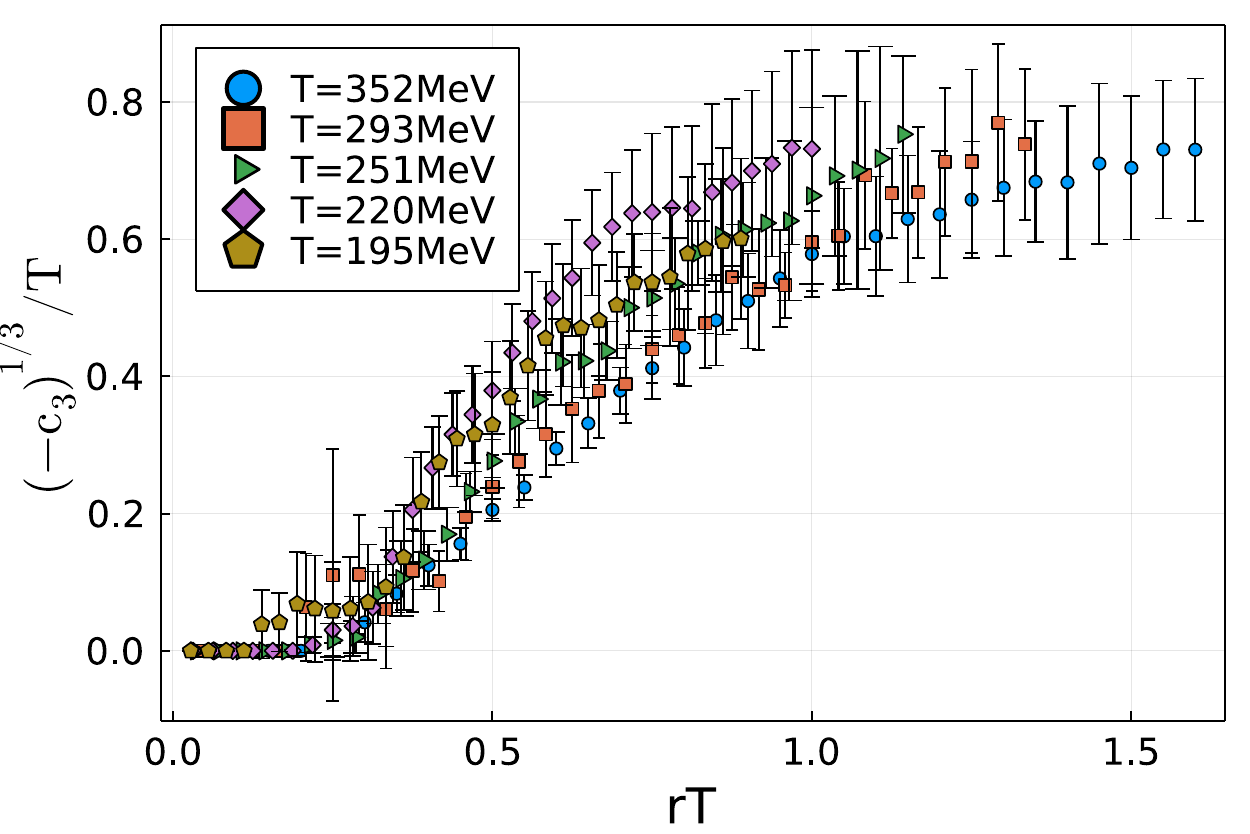}
\caption{$(-c_3)^{1/3}$ as a function of $r$ in temperature units for lattice spacing $a=0.0280$ fm
and different temperatures.}
\label{fig:c3_b8249}
\end{figure}

We also calculated the third cumulant of the spectral function using
our fitted spectral function based on the cut Lorentzian form. 
The result on $c_3$, which is the measure of skewness of the spectral
function, is shown in Fig. \ref{fig:c3_b8249}. We see that $-c_3$ is close
to zero at small $r$ but then rapidly increases with increasing $r$.
For very small distances, $r<5 a$ $\rho_r^{\rm low}(\omega,T)$ was not included
in the fit, and therefore, $c_3$ is exactly zero here.
Unfortunately, our lattice results are not precise enough 
to obtain $c_3$ using fits with Eq. (\ref{pol2}) extended to higher
order polynomials in the exponent. Thus at the present level of accuracy,
the short $\tau$ behavior of the effective masses can be parametrized 
solely by $m_1$ and $m_2$.

\newpage

\end{document}